# $Fe_3O_4$ Nano-octahedra/Vulcan XC72: Optimization and Combination with Solar-Based Electro-Fenton for Progestins Degradation

Juliana M. S. de Jesus[1*], Caroline de O. Carrilho[1], João P. C. Moura[1], Aline B. Trench[1], Caroline C. Augusto[1], Bruno L. Batista[1], Mauro C. dos Santos[1*]

[1]Laboratory of Electrochemistry and Nanostructured Materials (LEMN) - Center for Natural and Human Science (CCNH), Federal University of ABC (UFABC), CEP 09210-170, Avenida dos Estados, Bairro Bangu, Santo André, São Paulo, Brazil

*_Corresponding author._ E-mail address: julianams.silva@gmail.com

*_Corresponding author._ E-mail address: mauro.santos@ufabc.edu.br

## ABSTRACT

The widespread presence of synthetic progestins, such as levonorgestrel (LNG) and gestodene (GES), in aquatic environments poses significant ecotoxicological risks due to their endocrine-disrupting properties. In this study, nano-octahedral magnetite ($Fe_3O_4$-NO) was synthesized via a hydrothermal route and incorporated into gas diffusion electrodes (GDEs) supported on Vulcan XC72 to enhance the *in-situ* electrogeneration of hydrogen peroxide ($H_2O_2$). High-resolution Transmission Electron Microscopy, X-ray diffraction, SEM, X-ray photoelectron spectroscopy, and contact angle measurements thoroughly characterized the physicochemical and morphological properties of the materials. 3% $Fe_3O_4$-NO/C provided a 2-fold increase on $H_2O_2$ selectivity in comparison with Vulcan XC72. Electrochemical performance was optimized using a $2^3$ factorial design and principal component analysis (PCA), with current density, pH, and $Na_2SO_4$ concentration as variables. The optimized GDE (3% $Fe_3O_4$-NO/C) achieved a maximum $H_2O_2$ production of 0.44 ± 0.02 g $L^{-1}$ with a current efficiency of 43.1 ± 0.23% and a specific energy consumption of 0.012 ± 0.009 kWh $g^{-1}$. The electrode was further applied to the degradation of LNG and GES using solar and anodic-assisted electro-Fenton processes. Under optimal conditions, over 70 % removal of both progestins was achieved, with stable performance across three operational cycles. These findings demonstrate the potential of 3% $Fe_3O_4$-NO/C-GDEs as efficient, reusable cathodes for sustainable electrochemical advanced oxidation processes (EAOPs) in water treatment.

**KEYWORDS:** Hydrogen peroxide, Solar-based process, Electro-Fenton, Progestins.

**ABBREVIATIONS**

AOPs – Advanced Oxidation Processes
BDD – Boron-Doped Diamond
CE – Current Efficiency
DSA – Dimensionally Stable Anode
EDS – Energy-Dispersive X-ray Spectroscopy
EAOPs – Electrochemical Advanced Oxidation Processes
EF – Electro-Fenton
FFT – Fast Fourier Transform
FTIR – Fourier Transform Infrared Spectroscopy
GDE – Gas Diffusion Electrode
GES – Gestodene
$H_2O_2$ – Hydrogen Peroxide
HRTEM – High-Resolution Transmission Electron Microscopy
ICP-MS – Inductively Coupled Plasma Mass Spectrometry
LNG – Levonorgestrel
LSV – Linear Sweep Voltammetry
MQW – Milli-Q Water
ORR – Oxygen Reduction Reaction
PCA – Principal Component Analysis
PG – Polypropylene Glycol
PTFE – Polytetrafluoroethylene
RRDE – Rotating Ring–Disk Electrode
SEC – Specific Energy Consumption
SEM – Scanning Electron Microscopy
SP – Specific Production
STEM – Scanning Transmission Electron Microscopy
TEM – Transmission Electron Microscopy
TOC – Total Organic Carbon
UFLC – Ultra-Fast Liquid Chromatography
XPS – X-ray Photoelectron Spectroscopy
XRD – X-ray Diffraction

## 1. INTRODUCTION

Emerging contaminants in water, like synthetic progestins (LNG - levonorgestrel and GES - gestodene) used in contraceptives, pose risks to ecosystems and health. Even at trace levels (5-50 ng $L^{-1}$), they disrupt endocrine signaling, causing reproductive issues, fish feminization, and ecological imbalance. Conventional wastewater treatment often fails to remove these micropollutants, underscoring the need for advanced, affordable remediation technologies [1–5].

In response to the limited efficiency of conventional wastewater treatment processes for removing endocrine-disrupting micropollutants, advanced oxidation processes (AOPs) have emerged as promising alternatives for degrading chemically persistent compounds[6–10]. Among these technologies, electrochemical AOPs that generate hydrogen peroxide ($H_2O_2$) in situ via the two-electron oxygen reduction reaction ($2e^-$ ORR) (**Equations 1 – 3**) have gained particular attention because of their operational simplicity, environmental compatibility, and ease of integration with Fenton and photo-Fenton processes (**Eq. 4**)[10–14].

$$O_2 + 2H^+ + 2e^- \rightarrow H_2O_2 \qquad E^0 = +\,0.695 \text{ V vs. SHE} \qquad \textbf{(1)}$$

$$H_2O_2 + 2H^+ + 2e^- \rightarrow 2H_2O \qquad E^0 = +\,1.776 \text{ V vs. SHE} \qquad \textbf{(2)}$$

$$O_2 + 4H^+ + 4e^- \rightarrow 2H_2O \qquad E^0 = +\,1.229 \text{ V vs. SHE} \qquad \textbf{(3)}$$

$$Fe^{2+} + H_2O_2 \rightarrow Fe^{3+} + {}^{\bullet}OH + OH^- \qquad \textbf{(4)}$$

An important advantage of this approach is the elimination of external $H_2O_2$ dosing, which reduces safety concerns and operational costs [15–18]. Nevertheless, the practical implementation of electrochemical $H_2O_2$ production remains challenging, as high selectivity for the $2e^-$ ORR pathway must be achieved while minimizing parasitic reactions, energy losses, and peroxide decomposition under realistic operating conditions [15–17,19].

Within this framework, iron-based oxides, especially magnetite ($Fe_3O_4$), are increasingly seen as promising electrocatalysts for $H_2O_2$-related electrochemical applications. $Fe_3O_4$ features mixed-valence $Fe^{2+}/Fe^{3+}$ redox chemistry, chemical stability, affordability, and environmental safety, making it ideal for electro-Fenton systems [20–22]. Nanostructured $Fe_3O_4$ enhances oxygen activation and electron transfer and acts as an effective mediator for Fenton-type reactions through continuous redox cycling. Additionally, specific crystallographic facets on $Fe_3O_4$ nanostructures can affect the adsorption and transformation of ORR intermediates, potentially improving $H_2O_2$ selectivity[23]. Nonetheless, the limited electrical conductivity of iron oxides and their performance dependence on dispersion, accessible surface area, and interfacial charge transport underscore the need for conductive support [20–22].

Carbonaceous supports play a crucial role in overcoming these limitations, with Vulcan XC72 among the most widely used benchmark materials in electrochemical applications [24–28]. Vulcan XC72 offers high electrical conductivity, a mesoporous structure, and excellent chemical stability, enabling efficient electron transport and uniform dispersion of metal oxide nanoparticles. Its surface oxygen-containing functional groups promote strong interactions with iron oxide phases, improving catalyst anchoring and durability. When incorporated into gas diffusion electrodes (GDEs), Vulcan XC72 further enhances mass transport by facilitating the formation of an effective triple-phase boundary among gaseous oxygen, liquid electrolyte, and solid catalyst, an essential feature for efficient ORR-driven $H_2O_2$ electrogeneration [10,15,29].

Despite significant advances, the simultaneous optimization of catalyst composition, operational parameters, and process performance remains underexplored. Few studies have systematically evaluated the role of $Fe_3O_4$/C-based GDEs in the *in situ* generation of $H_2O_2$ and their application for the degradation of pharmaceutical contaminants, such as progestins [22,30,31].

In this sense, the present study aims**:** (i) to synthesize and characterize nano-octahedral $Fe_3O_4$ supported on Vulcan XC72 carbon; (ii) to evaluate its performance in gas diffusion electrodes for electrochemical $H_2O_2$ generation; (iii) to apply factorial design and principal component analysis (PCA) to optimize electrogeneration conditions; and (iv) to investigate the degradation efficiency of LNG and GES under different operational scenarios, including solar and anodic enhancement strategies.

## 2. MATERIALS AND METHODS

### 2.1 Synthesis of the Magnetite Nanooctahedra ($Fe_3O_4$ – NO)

All reagents used were of high purity, requiring no pre-purification, and were obtained from Sigma-Aldrich. The iron nanomaterials were synthesized using a hydrothermal method, adapted from Lei *et al.* (2017) [32]. Iron sulfate heptahydrate ($FeSO_4$ 7 $H_2O$, ≥ 99.0%, CAS 7782-63-0) was used as a metallic precursor and dissolved in 10 mL of ultrapure water (MQW, Millipore Mili-Q system, conductivity 18.2 MΩ cm) under magnetic stirring [23]. Subsequently, 22 mL of polypropylene glycol (PG, $C_3H_8O_2$, CAS 25322-69-4) was added to the slightly yellow solution, and the mixture continued to stir magnetically for one hour. 1 mmol $L^{-1}$ of sodium hydroxide (NaOH, ≥ 97.0%, CAS 1310-73-2) was added to reduce the iron, yielding a greenish solution. This mixture was subsequently transferred into a 70 mL stainless steel autoclave and heated in a muffle furnace for 20 hours at 200°C. Modifying the procedure was imperative, specifically by aerating nitrogen gas for 10 minutes [33]. Subsequently, following the hydrothermal process, all obtained nanomaterials underwent triple washing with a 50% v/v mixture of water and ethyl alcohol ($CH_3CH_2OH$, CAS 64-17-5) facilitated by centrifugation at 9000 rpm for 10 minutes. The supernatant from the washes was discarded, and the resulting pellet was dried in an oven at 80°C for 24 hours.

### 2.1.1 Electrocatalyst preparation

$Fe_3O_4$-NO and Vulcan XC72 carbon were the materials used in the present study, with $Fe_3O_4$-NO serving as a metallic source to investigate its improvement in addition to Vulcan XC-72, which was also used as a support matrix. $Fe_3O_4$-NO /C was prepared via wet impregnation, involving mixing 2.91 g of Vulcan XC72 carbon with 0.09 g of $Fe_3O_4$-NO (3% w/w nanostructure in the carbon matrix) in 100 mL of Milli-Q water. These mixtures were mechanically agitated for 5 hours and dried at 100 °C for 24 hours [34].

### 2.1.2 Gas-diffusion electrode (GDE) preparation

GDE 1 and 2, composed of Vulcan XC72 and 3% $Fe_3O_4$-NO/C, respectively, were prepared by mixing the electrocatalyst mass with 20% w/w of Polytetrafluoroethylene (PTFE dispersion, 60% w/w in water, CAS 9002-84-0) in 500 mL of Milli-Q water under constant stirring for 24 hours. Subsequently, the electrocatalyst was vacuum-filtered and dried at 100 °C for 4 hours to achieve a dry mass of 2–10%. The obtained material was hot-pressed (SOLAB SL-10/15-E) between two stainless steel plates (3.5 $cm^2$) and heated at 290°C, 4 tons for 2 hours [25,35].

### 2.1.3 Electrocatalyst characterization

High-resolution Transmission Electron Microscopy (HRTEM) was employed to assess the morphology, state of aggregation, and dimensions of the synthesized $Fe_3O_4$-NO. The microscope (JEOL JEM-2100) was used in conjunction with the Energy-dispersive X-ray Spectroscopy (EDS) module for semi-quantitative elemental analysis. The samplers (Formvar/Carbon-supported copper, 300 mesh, Merck) were prepared using a stereoscopic microscope (Digilab, DI-152 T). Additionally, the Scanning Electron Microscope (SEM FESEM JEOL JSM-7401F) was applied to obtain the structure distribution of $Fe_3O_4$-NO on carbonaceous support.

X-ray powder diffraction (XRD) measurements were performed in transmission geometry using a conventional D8 FOCUS diffractometer (Bruker AXS GmbH), operating at 40 kV and 40 mA, and equipped with a Ge(111) primary-beam monochromator, which provides $Cu_{K\alpha 1}$ radiation ($\lambda$ = 1.54056 Å). The samples were deposited between two sheets of cellulose acetate, and the sample holder was kept rotating during data collection. The integrated intensities were recorded with a linear detector model, Mythen 1 K (Dectris®, Baden, Switzerland), over 10°-90°, 2θ, at a scan rate of 2° $min^{-1}$.

Wettability was assessed using a contact angle meter (SEO Phoenix 300, Kromtek). 60 µL of nanomaterial suspension was placed on a glassy carbon plate. 10 µL of distilled water was deposited onto the nanoparticle films to measure contact angles. Surfaceware9 software captured the image and processed the data. Measurements were performed in triplicate.

X-ray photoelectron spectroscopy (XPS) analysis was performed using a Scienta Omicron $ESCA^+$ spectrometer with monochromatic Al Kα (1486.7 eV) radiation. Shirley's method improved accuracy and removed the inelastic background in the C 1s high-resolution core-level spectra. Spectral fitting was performed with CasaXPS software using unconstrained multiple-Voigt-profile fitting.

Elemental determinations were performed using an inductively coupled plasma mass spectrometer (ICP-MS, Agilent 7900, Hachioji, Japan). The certified reference material SRM 1643f (National Institute of Standards and Technology, USA) was employed for quality control. The instrumental operating conditions are summarized in **Table S1**.

## 2.2 Electrochemical measurements

### 2.2.1 Experimental design for *in-situ* $H_2O_2$ electrogeneration

Initially, a $2^3$ factorial design was employed to investigate the overall trends in *in-situ* $H_2O_2$ electrogeneration, to reduce energy consumption and increase current efficiency within a 60-minute process (**Table 1**). Three variables were analyzed at two levels: $X_1$ (current density – mA cm$^{-2}$); $X_2$ (pH), and $X_3$ ($[Na_2SO_4]_0$ – mol L$^{-1}$), composing eight factorial points (runs 1 to 8) and three replicates of the central point (runs 9 to 11) (**Table S2**).

$Na_2SO_4$ was used as the supporting electrolyte due to its high ionic conductivity and inertness over the tested potential range. Sulfate ions do not produce secondary oxidizing species or significantly complex iron, preserving oxygen reduction and Electro-Fenton processes. Alternatives such as chloride, nitrate, phosphate, or carbonate can cause parasitic reactions, radical scavenging, or iron complexation, thereby affecting hydrogen peroxide selectivity and reproducibility. Hence, $Na_2SO_4$ was chosen for reliable electrochemical control and comparison.

The independent variables were applied as a function of response: hydrogen peroxide electrogenerated (%) (**Eq. 5**), specific production of $H_2O_2$ (kWh $g^{-1}$) **(Eq. 6)**, and current efficiency (CE%) **(Eq. 7)** [36–38].

$$H_2O_2\ (\%) = \left(\frac{[H_2O_2]_t - [H_2O_2]_0}{[H_2O_2]_t}\right) \times 100 \quad \textbf{(5)}$$

$$SP_{H2O2}\ (kWh\ g^{-1}) = \frac{[H_2O_2] \times V}{E \times I \times t} \quad \textbf{(6)}$$

$$CE(\%) = \frac{2\ F\ [H_2O_2]\ V}{I\ t} \times 100 \quad \textbf{(7)}$$

Where 2 is the number of electrons exchanged for the reduction of $O_2$ to $H_2O_2$, F is the Faraday constant (96485 C $mol^{-1}$), [$H_2O_2$] is the hydrogen peroxide concentration (mol $L^{-1}$ (**Eq. 5**) and mg $L^{-1}$ (**Eq. 6**). V is the working volume (L), I is the applied electric current (A), E is the cell potential (V), and t is the electrolysis time (s)(**Eq. 6**) and h (**Eq. 7**).

**Table 1**. $2^3$ factorial design matrix with absolute and coded values and the responses relating to hydrogen peroxide production ($H_2O_2$), specific production of $H_2O_2$ ($SP_{H2O2}$), and current efficiency (CE) for the $H_2O_2$ electrogenerated by Vulcan XC72. Electrolysis time: 60 min.

| Runs | Coded Values | | | Original Variables | | | Responses | | | Extra Information |
|---|---|---|---|---|---|---|---|---|---|---|
| | $X_1$ | $X_2$ | $X_3$ | $j$ (mA cm$^{-2}$) | pH | $[Na_2SO_4]_0$ (mol L$^{-1}$) | $H_2O_2$ (%) | $SP_{H2O2}$ (kWh g$^{-1}$) | CE (%) | $[H_2O_2]_{60'}$ (g L$^{-1}$) |
| **1** | −1 | −1 | −1 | 25.0 | 3.0 | 0.1 | 88.5 | 0.092 | 52.2 | 0.099 |
| **2** | +1 | −1 | −1 | 250.0 | 3.0 | 0.1 | 89.5 | 0.007 | 5.90 | 0.109 |
| **3** | −1 | +1 | −1 | 25.0 | 9.0 | 0.1 | 88.0 | 0.163 | 50.1 | 0.095 |
| **4** | +1 | +1 | −1 | 250.0 | 9.0 | 0.1 | 92.8 | 0.005 | 8.50 | 0.158 |
| **5** | −1 | −1 | +1 | 25.0 | 3.0 | 1.0 | 90.9 | 0.173 | 65.9 | 0.125 |
| **6** | +1 | −1 | +1 | 250.0 | 3.0 | 1.0 | 94.8 | 0.013 | 11.9 | 0.221 |
| **7** | −1 | +1 | +1 | 25.0 | 9.0 | 1.0 | 92.0 | 0.182 | 75.1 | 0.143 |
| **8** | +1 | +1 | +1 | 250.0 | 9.0 | 1.0 | 95.1 | 0.012 | 12.7 | 0.236 |
| **9** | 0 | 0 | 0 | 137.5 | 6.0 | 0.55 | 95.2 | 0.026 | 23.3 | 0.241 |
| **10** | 0 | 0 | 0 | 137.5 | 6.0 | 0.55 | 95.1 | 0.026 | 22.5 | 0.233 |
| **11** | 0 | 0 | 0 | 137.5 | 6.0 | 0.55 | 95.9 | 0.033 | 27.1 | 0.281 |

The pH values in **Table 1** were initially adjusted with 0.01 mol L$^{-1}$ NaOH or 0.01 mol L$^{-1}$ $H_2SO_4$. The electrogenerated hydrogen peroxide was determined by spectrophotometry using the ammonium metavanadate [39] (60.0 mmol L$^{-1}$ with 0.58 mol L$^{-1}$ of $H_2SO_4$) and the ammonium molybdate [40] (2.4 mmol L$^{-1}$ with 0.5 mol L$^{-1}$ of $H_2SO_4$) method due to the ferric interference in the first method (**Figure S1**).

Subsequently, to investigate the effect of 3% $Fe_3O_4$-NO/C in *in-situ* $H_2O_2$ electrogeneration, the optimal operational conditions for $H_2O_2$ accumulation (%) were applied to the modified GDE and rotating ring-disk electrode (RRDE) measurements.

### 2.2.2 Principal component analysis (PCA)

PCA was applied as an exploratory multivariate statistical tool to evaluate the impact of operational parameters on electrochemical performance. This methodology decreases the dimensionality of the dataset by converting correlated variables into a reduced set of orthogonal components (principal components) that preserve most of the original variance [41]. The PCA was implemented on the experimental dataset acquired from a $2^3$ factorial design (**Table 1**), encompassing independent variables (**Table S2**). The initial three principal components (PC1, PC2, and PC3) collectively explained 100% of the variance, allowing for the visualization of a three-dimensional data structure. Score plots revealed distinct clustering patterns associated with each response variable, whereas loading vectors illustrated the contributions and directional influences of the operational parameters on system behavior. This analysis helped identify key experimental factors that support performance metrics and strengthened the interpretation of interaction effects inherent to the factorial design [41].

### 2.2.3 RRDE measurements

RDE experiments were performed at rotation rates ranging from 100 to 1600 rpm, using a potentiostat/galvanostat (Autolab PGSTAT 302 N). The working electrode consisted of a glassy carbon disk (0.2475 $cm^2$) surrounded by a platinum ring (0.1866 $cm^2$), with a collection efficiency (N) of 0.21[42]. A platinum plate and Ag/AgCl electrode served as the counter and reference electrodes, respectively. The supporting electrolyte was 1 mol $L^{-1}$ $Na_2SO_4$ at pH 6.0.

The electrocatalyst inks were prepared by dispersing 2 mg mL-1 catalyst in deionized water under ultrasonic agitation, then depositing 20 µL onto a glassy carbon disk. After drying at ambient conditions, a 20 µL solution of Nafion in deionized water (1:100 v/v) was applied to the catalyst layer and thoroughly dried. Before testing, the electrolyte was purged with nitrogen for 30 min to activate it electrochemically and then kept under continuous oxygen flow during measurements.

Linear sweep voltammetry (LSV) was carried out on the disk electrode in the potential range from 0 V to −1.0 V vs. Ag/AgCl at a negative scan rate, while the platinum ring was held at 0.8 V vs. Ag/AgCl to oxidize any $HO_2^-$ species reaching the ring to $O_2$. The number of electrons transferred in the ORR and the selectivity toward $H_2O_2$ production were determined using **Equations (8)** and **(9)**:

$$\%H_2O_2 = \frac{\frac{2I_r}{N}}{I_d + \frac{I_r}{N}} \times 100 \quad \textbf{(8)}$$

$$n = 2\,[(X_{H2O2}) + 1] \quad \textbf{(9)}$$

*Ir*, *Id*, and *N* represent the ring current, disk current, and RRDE collection efficiency, respectively [42,43].

### 2.3 Electrodegradation of LNG and GES

The feasibility of *in-situ* $H_2O_2$ electrogeneration to degrade LNG and GES with 3% $Fe_3O_4$-NO/C was investigated by applying the optimized conditions of previous assays ($j$ = 137.5 – 250.0 mA $cm^{-2}$; pH = 6.0 – 9.0 and $[Na_2SO_4]_0$ = 1.0 mol $L^{-1}$). Although the oxygen reduction reactions involve proton consumption and generation, buffer solutions were not used in the optimization studies. This choice was made to avoid potential interference from buffering species, which may act as hydroxyl radical scavengers or complex iron species, thereby altering both $H_2O_2$ electrogeneration and Electro-Fenton reactivity. Instead, the electrolyte's initial pH was carefully adjusted before electrolysis and monitored during operation, allowing the system to evolve under conditions more closely resembling practical Electro-Fenton applications. This approach ensures that the observed effects of pH, current density, and electrolyte concentration reflect intrinsic electrochemical behavior rather than buffer-induced artifacts.

In this sense, control experiments (**Table 3**) were conducted using different anodes (Pt, DSA, and BDD) to investigate the effect of anodic oxidation and solar light (Solar simulator; working distance of 8.0 cm = 100 mW $cm^{-2}$; Oriel® LCS100™ with AM1.5G filter, Newport) combined with $H_2O_2$ electrogenerated (**Figure 1**).

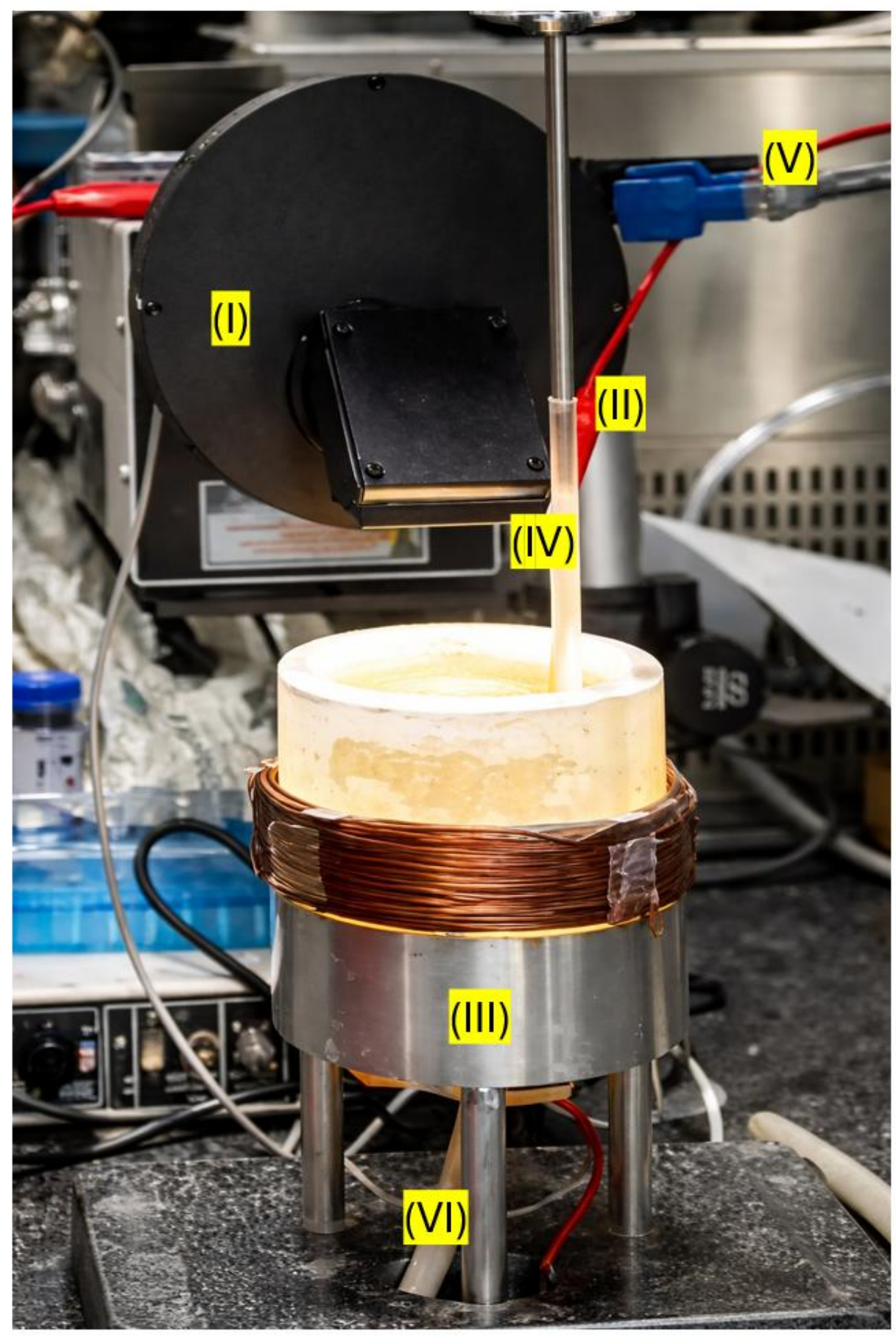

**Figure 1.** Representation of the experimental setup used for electrochemical experiments: (I) Solar simulator; (II) Anode (Pt, BDD, or DSA); (III) GDE; (IV) mechanical stirring; (V) Power supply, and (VI) oxygen supply.

GES and LNG concentrations were analyzed using ultra-fast liquid chromatography (UFLC) with Shimadzu equipment (LC-20AD), which included a UV–Visible detector (SPD-20A) and a C18 column (ACE, 250 mm × 4.60 mm, 5 µm). An isocratic method was employed, utilizing a mobile phase of 70% methanol and 30% water with 1% v/v acetic acid. Detection of both hormones occurred at 244 nm. The parameters set for the analysis were a 20 µL injection volume, an oven temperature of 40 °C, and a flow rate of 1.0 mL $min^{-1}$. Under these settings, GES and LNG showed retention times of 8.0 and 10.0 min, respectively. **Table S3** presents the validation parameters of the calibration curves. Additionally, total organic carbon (TOC) removal was determined using a Shimadzu TOC-VCPN before and after the electrochemical process.

Additionally, to evaluate the reusability of GDE 2 (3% $Fe_3O_4$-NO/C) after the electrogeneration and degradation experiments, three 60-minute cycles of the Solar Electro-Fenton process were conducted with LNG and GES solution. Energetic and removal parameters were obtained, along with total Fe measurements, to assess the leaching of GDE into the treated solution.

# 3. RESULTS AND DISCUSSION

## 3.1 Morphological, chemical, and structural $Fe_3O_4$ and $Fe_3O_4$-NO/C characterization

TEM analysis revealed that the synthesized $Fe_3O_4$ nanoparticles exhibit well-defined nano-octahedral and quasi-cubic morphologies (**Figure 2a)** with particle sizes ranging from 50 to 100 nm (**Figure 2c**), and no synthesis contamination traces (**Figure 2b**). These faceted structures indicate controlled crystal growth along specific crystallographic directions, characteristic of the inverse spinel structure of magnetite. Fast Fourier Transform (FFT) analysis performed on high-resolution TEM images confirmed the high crystallinity of individual nanoparticles (**Figure 2d**). The FFT pattern displayed distinct diffraction spots corresponding to spatial frequencies of 3.90, 4.23, and 5.03 $nm^{-1}$, which are associated with interplanar spacings of 0.256, 0.236, and 0.199 nm, respectively. These values match well with the (311), (400), and (511) planes of the cubic $Fe_3O_4$ phase (JCPDS No. 19-0629). The symmetry and clarity of the diffraction pattern suggest that the analyzed crystal was oriented along a [011] or [111] zone axis, corroborating the uniform morphology observed in the TEM images. Together, the TEM and FFT analyses confirm the formation of single-crystalline $Fe_3O_4$ nanoparticles with well-resolved crystallographic domains.

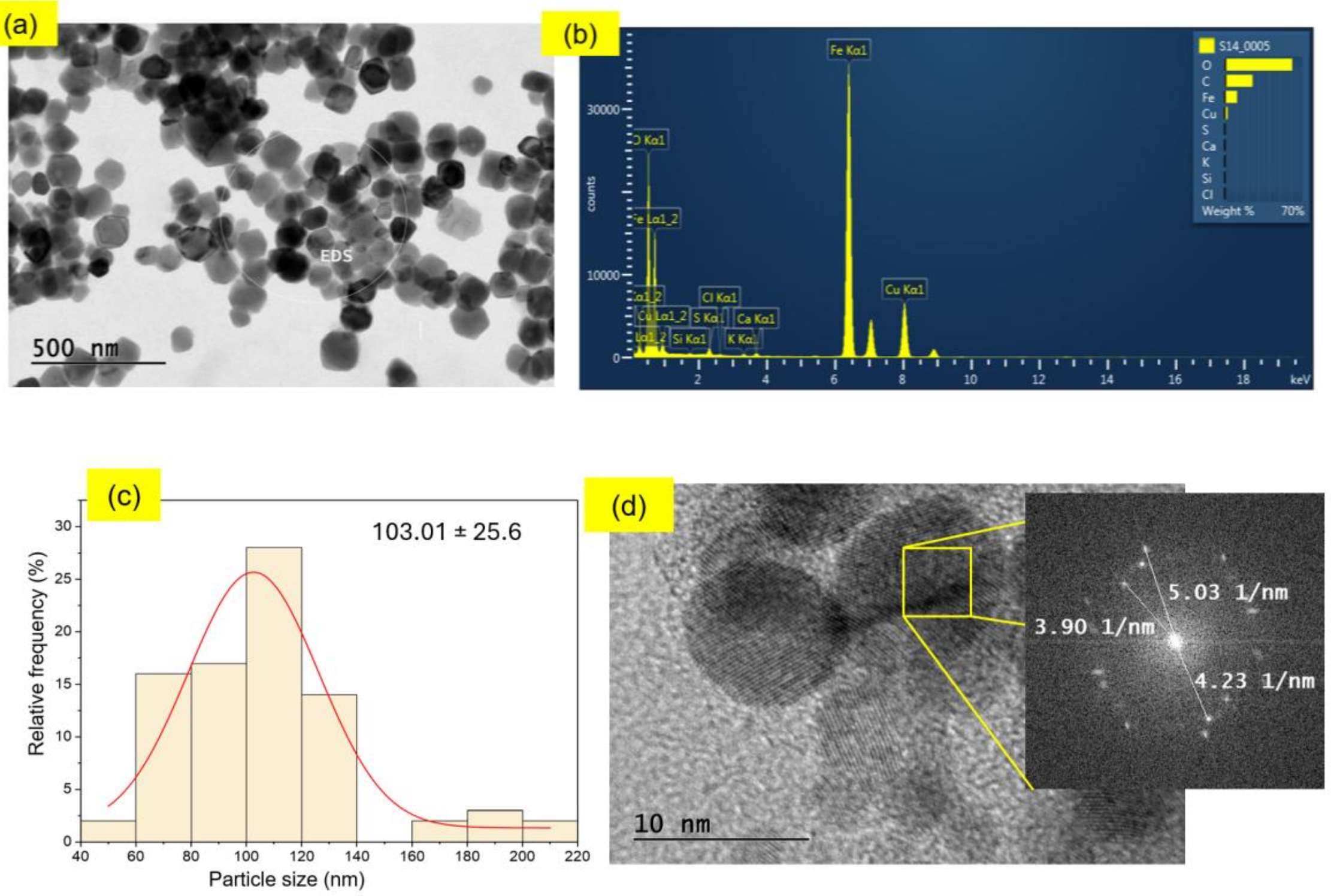

**Figure 2.** (a) TEM micrograph of the $Fe_3O_4$ nano-octahedra. Additional results: (b) EDS spectrum of the $Fe_3O_4$; (c) histogram of the particle size distribution of $Fe_3O_4$ nano-octahedra and FFT (d) of the region indicated by the yellow square of the $Fe_3O_4$.

XRD analysis (**Figure 3**) confirmed the formation of crystalline and phase-pure $Fe_3O_4$ nanoparticles, with diffraction peaks at $2\theta \approx 30.1°$, 35.5°, 37.1°, 43.1°, 53.5°, 57.0°, 62.6°, and 74.0°, indexed to the (220), (311), (222), (400), (422), (511), (440), and (533) planes of the cubic inverse spinel structure (JCPDS No. 19-0629). The intense (311) reflection, consistent with FFT analysis of HRTEM images (**Figure 2d**), indicates preferential crystallographic orientation. The crystallite size (d), estimated from the full width at half maximum (FWHM) of the (311) peak at 35.6° using the Scherrer equation (**Eq. S1**)[44], was 70 nm, suggesting the formation of large, well-crystallized spinel domains. Similar values have been reported for magnetite nano-octahedral particles synthesized via hydrothermal and coprecipitation methods (50–80 nm)[45,46]. Despite the Scherrer equation providing only the coherent diffraction domain size, which may differ from the particle size observed by HRTEM (**Figure2**) due to aggregation or polycrystallinity, the calculated value agrees with the narrow XRD peaks, corroborating the high crystallinity of the $Fe_3O_4$ phase.

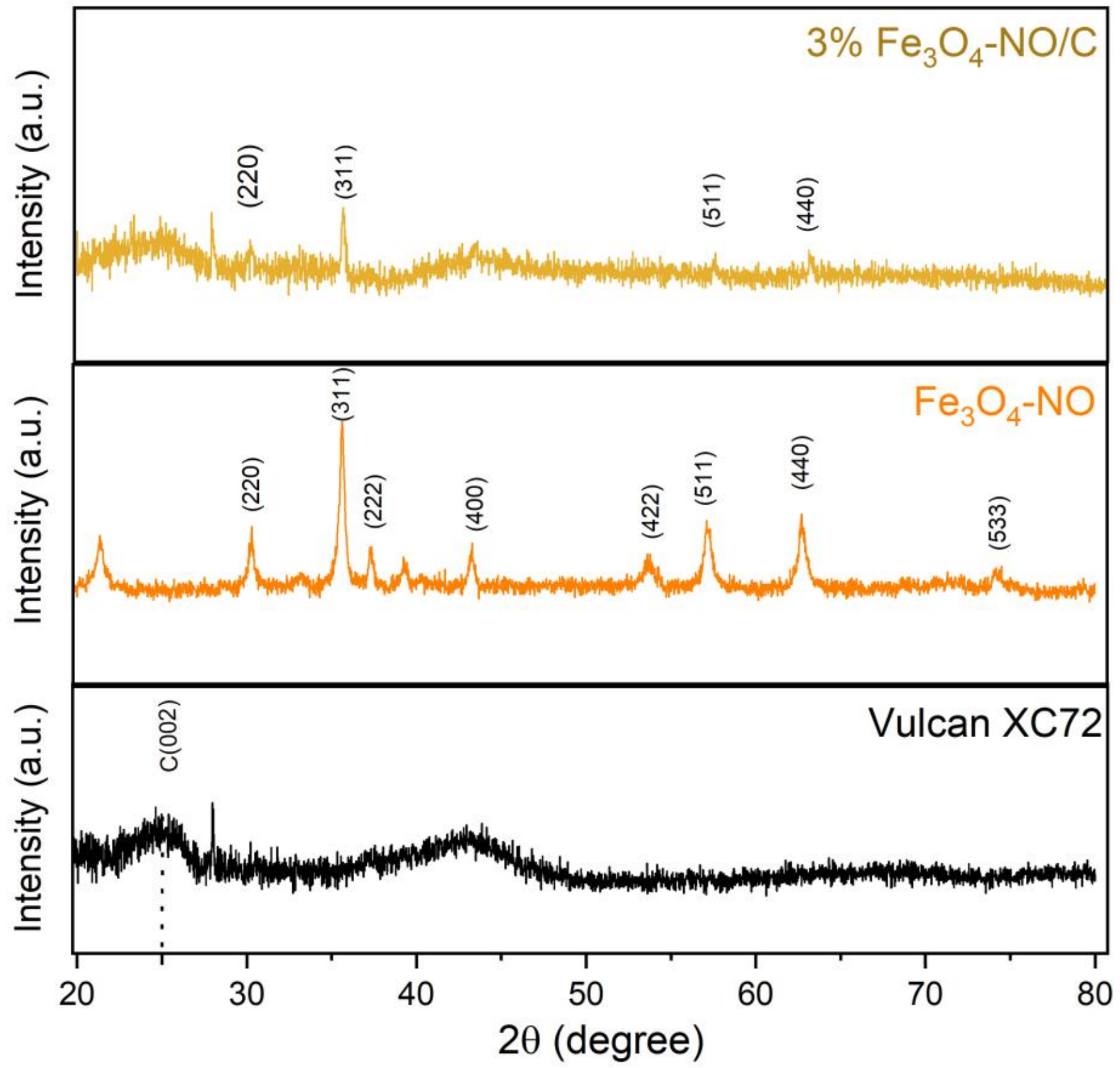

**Figure 3**. X-ray diffraction patterns of the powders applied for GDE preparation.

In the 3% $Fe_3O_4$-NO/C electrocatalyst pattern, the characteristic $Fe_3O_4$ peaks remain visible, albeit with reduced intensity due to the lower loading and high dispersion on the carbon matrix. The Vulcan XC72 support exhibits a broad, low-intensity peak near 25°, attributed to the (002) plane of amorphous carbon, confirming its low crystallinity.

The preservation of the magnetite diffraction features in the composite confirms the structural integrity of $Fe_3O_4$ after carbon incorporation.

SEM analysis of the 3% $Fe_3O_4$-NO/C electrocatalyst (**Figure 4**) revealed a hierarchically structured agglomerate composed of nanoscale $Fe_3O_4$ particles dispersed over the carbonaceous matrix. The micrograph, obtained at a magnification of 3,500×, shows the formation of densely packed clusters with a rough surface texture, indicating the presence of high-surface-area features. Distinct nano-octahedral crystals are visible on the surface, consistent with the faceted morphologies observed in the TEM analysis. These well-defined polyhedral particles are homogeneously distributed within the agglomerates, suggesting successful incorporation and partial exposure of $Fe_3O_4$ on the conductive carbon support. Typical aggregation was observed for nanoparticle–carbon composites, which can enhance electrical conductivity and catalytic contact in electrocatalytic applications. Overall, the SEM results corroborate the effective dispersion and morphological preservation of $Fe_3O_4$-NO within the carbon matrix.

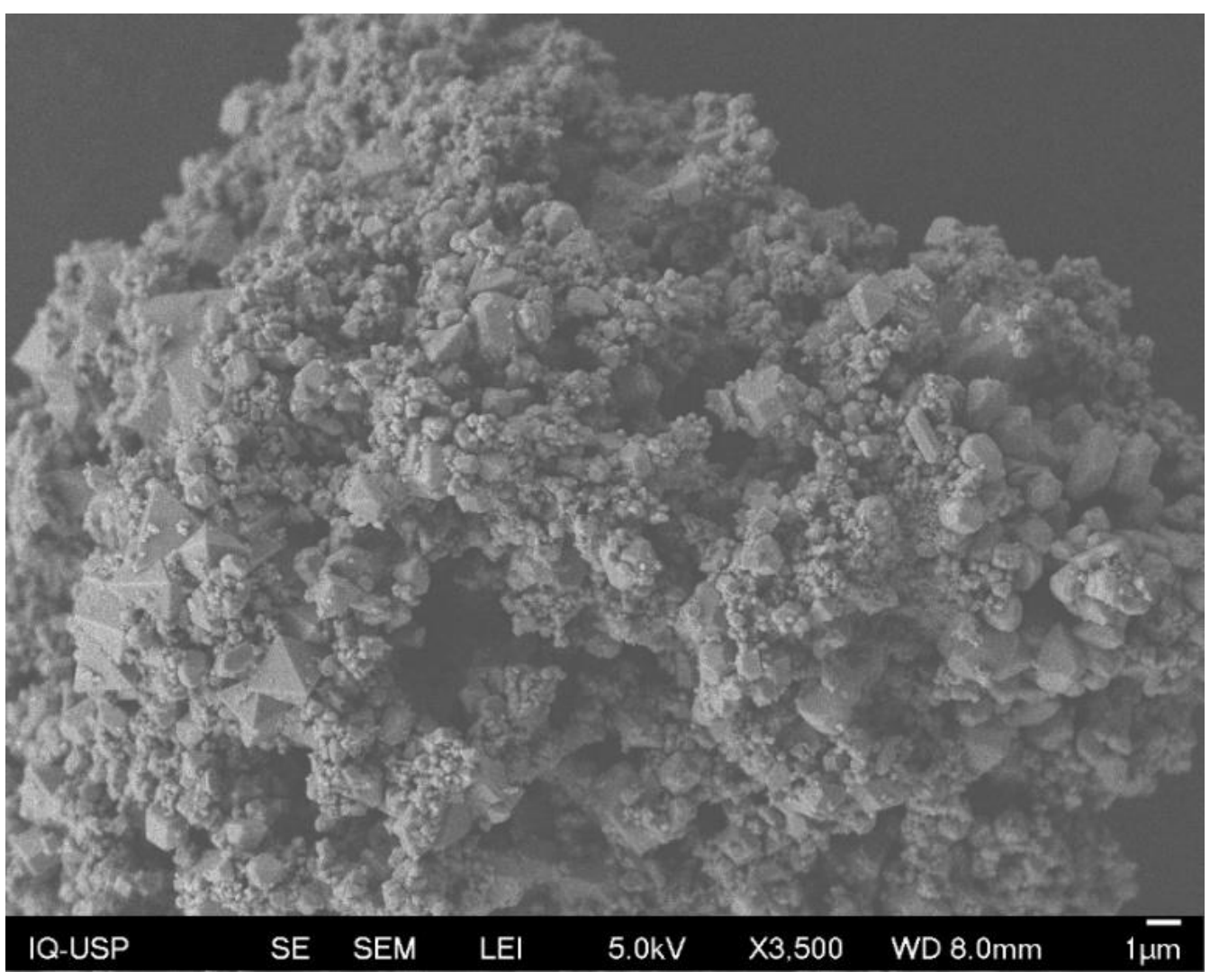

**Figure 4**. Image of backscattered electrons obtained by SEM of the 3% $Fe_3O_4$-NO/C electrocatalyst applied on GDE 2.

**Figure 5** shows that the wettability of Vulcan XC72 decreases by a factor of 1.6, particularly at a 3% loading, thereby improving surface hydrophilicity. This can enhance electrocatalytic performance by promoting better triple-phase boundary formation, crucial in $H_2O_2$ electrogeneration and ORR systems [35,47].

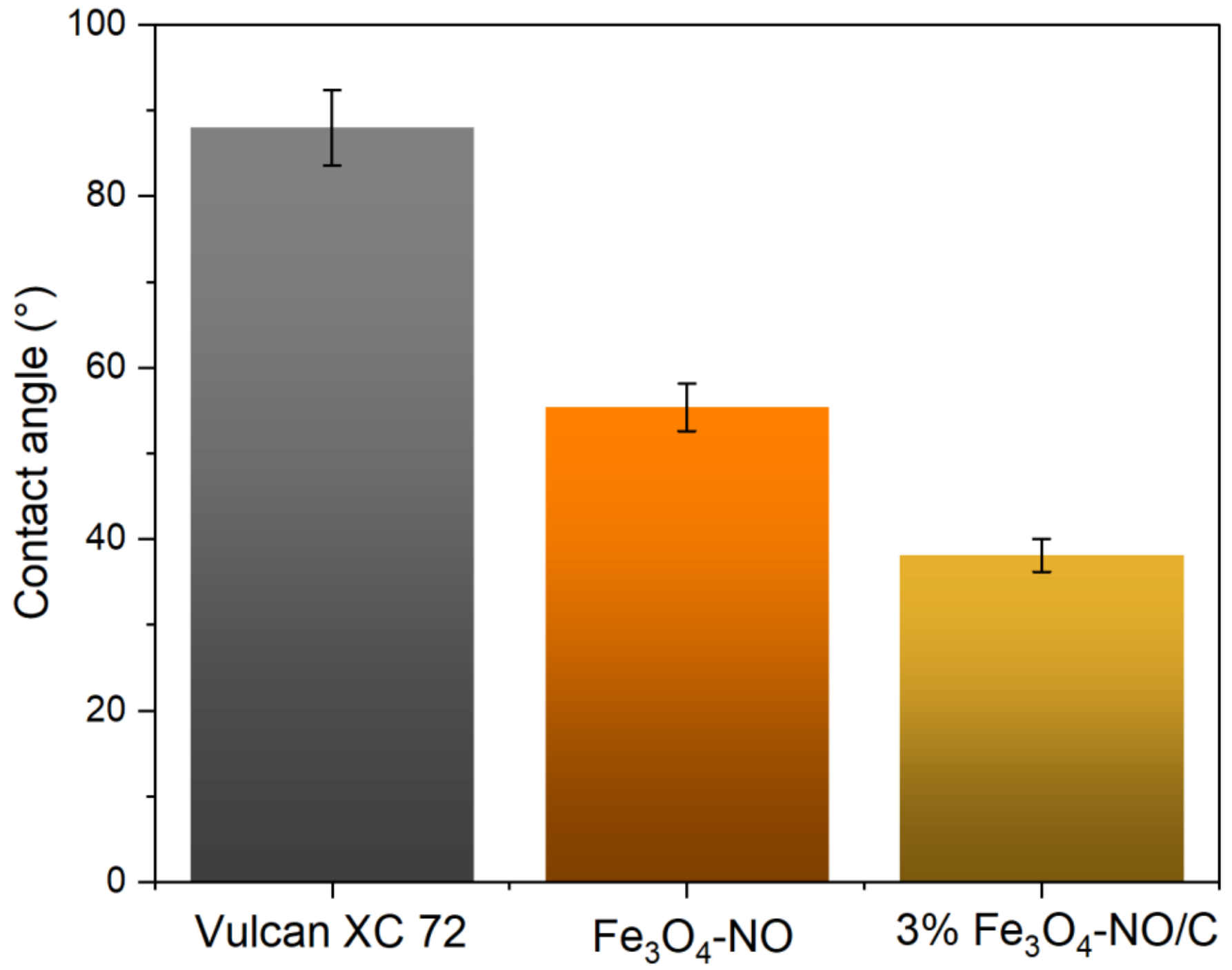

**Figure 5.** Contact angle of the electrocatalysts studied. (n = 3)

XPS analysis of the $Fe_3O_4$-NO sample was performed to confirm its chemical composition and identify the surface electronic states (**Figure 6**). **Figure 6a** shows the XPS spectrum of Fe 2p for $Fe_3O_4$-NO. The binding energy peak around 711 eV was assigned to the Fe $2p_{3/2}$ level, while the signal near 724 eV corresponds to the Fe $2p_{1/2}$ level. This spectrum was deconvolved into two spin-orbit doublets, corresponding to $Fe^{3+}$ and $Fe^{2+}$ surface contributions, plus two satellites [48,49].

The XPS spectrum of O 1s was also investigated (**Figure 6b),** which was deconvolved into two components: the 530 eV component was attributed to $O^{2-}$ ions in the iron oxide lattice, and the 532 eV component was attributed to surface OH groups or oxygen vacancies in $Fe_3O_4$ [12].

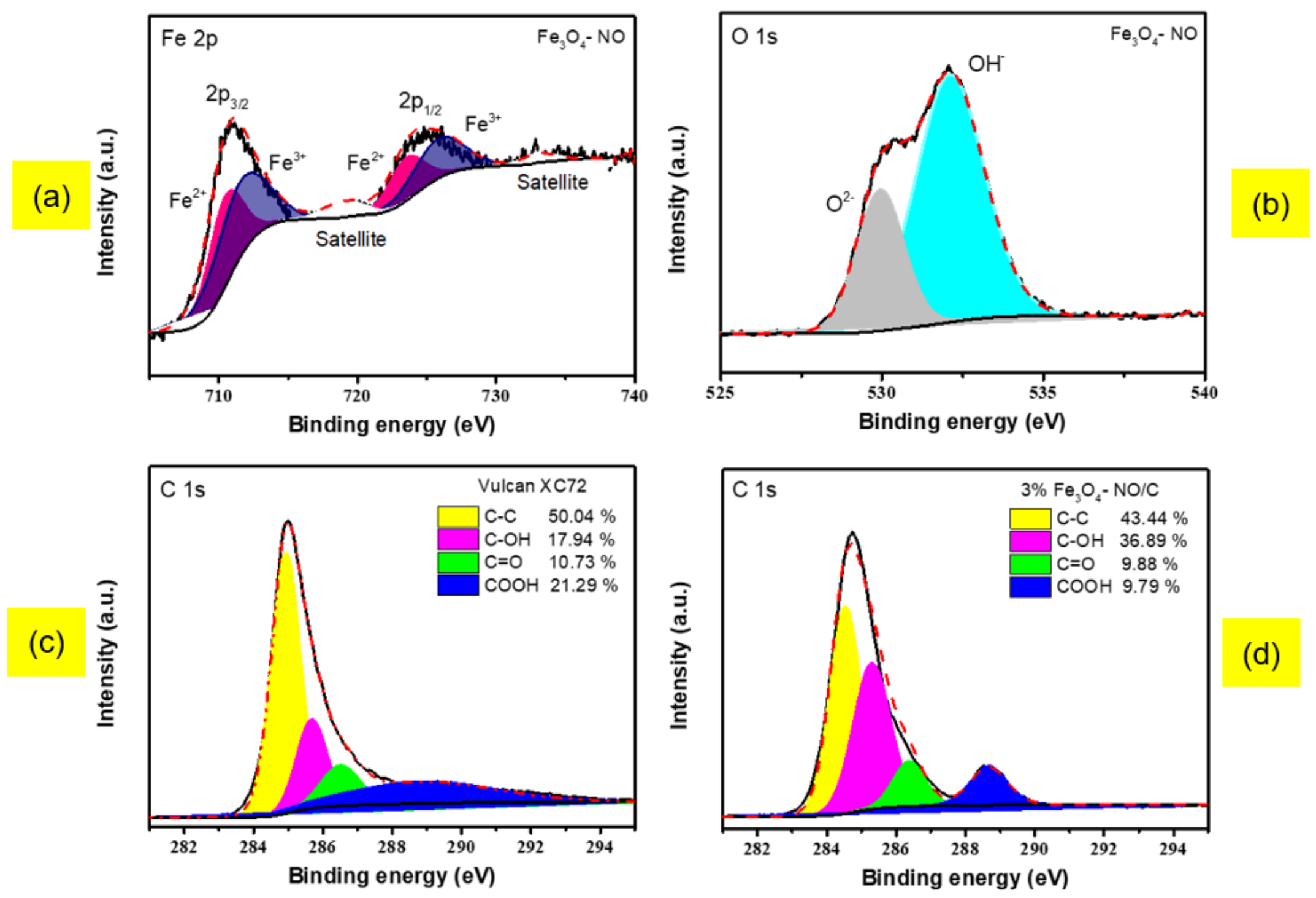

**Figure 6**. High-resolution spectra of (a)Fe 2p, (b) O 1s for $Fe_3O_4$-NO; (c) C 1s for Vulcan XC72; (d) 3% $Fe_3O_4$-NO/C.

**Figure 6c** revealed significant changes in the surface chemistry of the Vulcan XC72 carbon following the incorporation of 3% $Fe_3O_4$-NO/C. **Figure 6c** was deconvoluted using four components corresponding to the C–C, C–OH, C=O, and COOH bonds, located at binding energies of approximately 284 eV, 285, 286, and 289 eV, respectively [50,51]. For the Vulcan XC72, the dominant contribution was attributed to graphitic C–C bonds (50.04%), followed by oxygenated functionalities including C–OH (17.94%), C=O (10.73%), and COOH (21.29%).

After $Fe_3O_4$ incorporation (**Figure 6d**), a noticeable decrease in C–C content (43.44%) and COOH groups (9.79%) was observed, suggesting partial oxidation and possible consumption of carboxylic sites during nanoparticle anchoring. Conversely, the C–OH content increased markedly to 36.89%, indicating a higher degree of hydroxylation, which enhanced hydrophilicity and facilitated electron transfer processes, as observed in the wettability results (**Figure 5**). These surface modifications reflect the successful functionalization of the carbon matrix and the effective interaction between $Fe_3O_4$ nanoparticles and the carbon support, potentially improving the material's electrocatalytic properties [25,45].

## 3.2 Electrochemical assessment

### 3.2.1 Optimization of *in-situ* $H_2O_2$ electrogenerated

**Table 1** presents the $2^3$-factorial design for statistical analysis performed in Statistica®, including the dependent variables (**Table S2**) applied as responses to investigate the main factors contributing to hydrogen peroxide electrogeneration. Regarding the data (**Table 1**), the responses achieved maximum performance in different experimental conditions: (i) highest hydrogen peroxide production (95.9% = 0.281 g $L^{-1}$ in 60 minutes of electrolysis (**Fig. S2**), run 11: $X_1$ = 137.5 mA $cm^{-2}$; $X_2$ = 6.0 and $X_3$ = 0.55 mol $L^{-1}$ ; (ii) lowest specific production (0.005 kWh $g^{-1}$, run 4: $X_1$ = 250 mA $cm^{-2}$; $X_2$ = 9.0 and $X_3$ = 0.1 mol $L^{-1}$); and highest current efficiency (75.1%, run 7: $X_1$ = 25 mA $cm^{-2}$; $X_2$ = 9.0 and $X_3$ = 1.0 mol $L^{-1}$). These results highlight the complexity of electrochemical $H_2O_2$ generation on carbon-based materials such as Vulcan XC72, where achieving higher current efficiency (CE) and lower specific energy consumption (SEC) remains a key challenge.

Another piece of evidence underscoring the complexity of hydrogen peroxide electrogeneration is provided by the mathematical model derived from **Tables S4-S6**. The coefficient of determination for the three models (**Eqs. 10-12**) indicates that 96.5%, 99.7%, and 99.5% of the data are explained by the models at the 95% confidence level. The impact of more than two independent variables was assessed utilizing ANOVA (**Tables S7 - S9**), for which the significance level (α) is set at 5% ($p < 0.05$).

**Eqs. 10-12** show the distinct effects of the dependent variables on the responses. For example, **Eq. 10** did not present interaction effects among the variables, as observed in **Eq. 11**, suggesting a simple process description. However, there was significant curvature ($p < 0.05$) in **Eqs. 10-12**, indicating that the linear model obtained did not account for all the factors contributing to *in-situ* $H_2O_2$ electrosynthesis.

$$H_2O_2\ (\%) = 91.45 + 1.60\ X_1 + 1.75\ X_3 + 3.95\ X^2 \quad \textbf{(10)}$$

$$SP_{H2O2}\ (\text{kWh g}^{-1}) = 0.081 - 0.072\ X_1 + 0.01\ X_2 + 0.01\ X_3 - 0.053\ X^2 - 0.01X_1X_2 - 0.01X_1X_3 - 0.01X_2X_3 \quad \textbf{(11)}$$

$$CE\ (\%) = 34.95 - 25.86\ X_1 + 6.43X_3 - 10.66\ X^2 \quad \textbf{(12)}$$

Based on this, the factorial design highlights the significant role of operational parameters:

(i) Effect of Current Density: The current density exhibited a beneficial effect on the electrogeneration of hydrogen peroxide (%), while simultaneously producing adverse effects on both energy responses (SP and CE). As shown in the mean plots (**Figure S3a and a'**), increasing current density from the low (-1) to high (+1) level resulted in a substantial reduction in SP, from approximately 0.15 to nearly zero (kWh $g^1$)—indicating a more energy-efficient process at higher current densities. However, this increase in energy efficiency was accompanied by a marked decline in CE, which dropped from around 62% to below 10%. This inverse relationship suggests that although higher current densities accelerate the production of oxidants, they also promote parasitic side reactions (e.g., $4e^-$ ORR or $H_2O_2$ decomposition), thereby reducing selectivity toward $H_2O_2$. Therefore, it demonstrated interaction effects with pH and electrolyte support concentration (**Eq. 11**). This outcome signifies an antagonistic relationship in the electrochemical process; specifically, to achieve optimal $H_2O_2$% production, it is necessary to utilize elevated current density levels, which is related to enhanced electron transfer rates at the electrode/solution interface, promoting the 2-electron oxygen reduction pathway to $H_2O_2$ [22]. Nevertheless, the SP and CE will be diminished, which should be avoided from an energy and cost-efficiency standpoint. Felisardo *et al.* [52] investigated the application of a GDE/PL6C electrode in $Na_2SO_4$ solution (0.3 mol $L^{-1}$, pH 5.26). Electrolysis was carried out at current densities of 25, 50, 75, and 100 mA $cm^{-2}$ for 100 min. Results showed a direct proportionality between current density and $H_2O_2$ production. Doubling the current from 25 to 50 mA $cm^2$ nearly doubled the $H_2O_2$ concentration, from 200 to 400 mg $L^{-1}$. The 75 and 100 mA $cm^2$ increases led to approximately 3- and 4-fold increases, from 700 to 945 mg $L^1$. In contrast to the findings of Felizardo *et al.* (2025)[52], the improvement in $H_2O_2$ production can also be limited by the supporting electrolyte concentration, which exhibits an intrinsic and adverse effect.

(i) Effect of pH: Previous studies have shown that the electrolyte pH highly influences the selectivity and activity of $2e^-$ ORR catalysts [24,26]. Therefore, understanding the pH effect is essential for designing efficient catalysts tailored to different electrolyte environments. The obtained models (**Eqs. 10 and 11**) highlight the adverse impact of the pH on the electrochemical process. Regarding $H_2O_2$ electrogeneration, the pH should be kept low (pH $\leq$ 6.0), whereas for SP improvement, it should be kept high (pH $\geq$ 6.0) (**Figure S3b**). This numerical difference also indicates a mechanistic distinction. Carbon-based catalysts exhibit weaker *OOH interactions, promoting disruption of intermediates and favoring $H_2O_2$ formation in acidic conditions (**Eq. 4**). Theoretical and experimental studies confirm that acidic environments enhance selectivity for $H_2O_2$. In contrast, alkaline conditions shift the pathway toward the formation of the unstable hydroperoxide ion ($HO_2^-$) (**Eq. 3**) [53].

(ii) Effect of $[Na_2SO_4]_0$: The concentration of the support electrolyte plays a role in the electrochemical process, essentially in *in-situ* $H_2O_2$ electro generation. In this last application, the type and concentration of the electrochemical solution affect the occurrence of side reactions, thereby reducing ohmic drop, improving electrolyte conductivity, and enhancing electrode stability. The effect of electrolyte concentration was evaluated, looking for the compression of its impact over the proposed variables, regarding the models obtained (**Eqs. 10 to 12**), all of which showed a significant effect of $X_3$ ($[Na_2SO_4]_0$). Nonetheless, the model helps us assess the operational tendency for this parameter, which should be high to increase energy efficiency **(Figure S2)**.

A PCA was applied to elucidate the correlation among the responses obtained to address the operational roles for *in-situ* electrogeneration. **Figure 7a** shows the distribution of experimental runs according to three principal components (PC1, PC2, and PC3), which explain 99.5% of the total variance (67.5%, 32.0%, and 0.6%, respectively). Experimental conditions were varied according to a $2^3$ factorial design (**Table 1**), and the responses evaluated were current efficiency (CE, blue), hydrogen peroxide generation ($H_2O_2$, red), and specific energy consumption (SP, yellow). As defined by the blue and yellow ellipsoids (**Figure 7a),** SP and CE merge, indicating a more structured and predictive model of the experimental variables. They are also directly affected by pH, current density, and $[Na_2SO_4]_0$, which are indicated by the loadings (blue arrows). The alignment and magnitude of the loadings reflect their influence on the variance structure: (i) $[Na_2SO_4]_0$ is strongly associated with PC2 (SP); (ii) pH and current density are positively impacting PC1; and $H_2O_2$ yield seems to have low alignment with any single variable, indicating a more complex dependence on the studied parameters.

To further elucidate sample grouping patterns, a two-dimensional PCA score plot (PC1 vs. PC2) was also constructed **(Figure 7b)**. This visualization confirmed the clustering tendencies observed in the 3D model **(Figure 7a)**, where experimental conditions led to well-defined response groups. The score plots distinguished samples with low $H_2O_2$ yield ($H_2O_2$ %, represented by black squares), which are clearly separated from those with high current efficiency and specific production (CE and SP, represented by blue triangles and red circles, respectively), based on their projection onto PC1 **(Figure 7b)**. Overlapping ellipses for CE and SP responses suggested that these two metrics were simultaneously favored under similar operational conditions, consistent with the directionality of the loadings observed in the multivariate space **(Figure 7a)**.

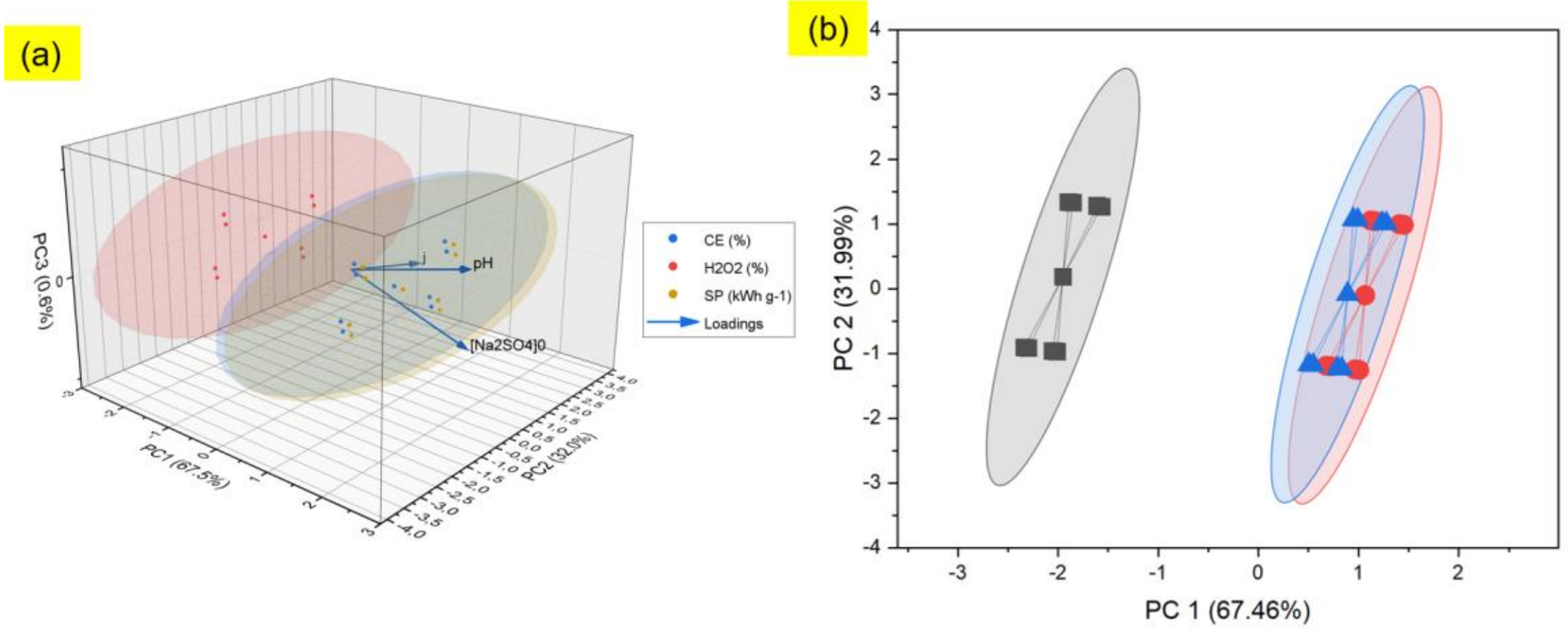

**Figure 7**. Loading plots (a) and scores (b) from PCA of electrochemical performance.

Based on this, the multivariate approach reveals grouping patterns among the responses, highlighting the optimal ranges of operational parameters for energetic factors (CE and $SP_{H2O2}$) and $H_2O_2$% (**Table S10**). According to the results, the investigation on *in-situ* $H_2O_2$ electrogeneration with GDE 2 and the combined process for progestin degradation will be conducted at $[Na_2SO_4]_0$ = 1.0 mol $L^{-1}$, pH = 6.0, and $j$ = 137.5 mA $cm^{-2}$.

3.2.2 RRDE, n, and $\%H_2O_2$

RRDE measurements assessed the ORR performance for Vulcan XC72 and 3% $Fe_3O_4NO/C$ in $O_2$-saturated conditions and 1.0 mol $L^{-1}$ $Na_2SO_4$ at pH = 6.0. **Figure 8a** presents the RRDE voltammograms obtained at pH 6.0, which reveal distinct electrocatalytic behaviors for Vulcan XC72 and 3% $Fe_3O_4NO/C$, confirming the contrasting reaction pathways and kinetic regimes governing the ORR. Vulcan XC72 exhibited lower disk currents (-190.0 to -390.0 μA) across all rotation rates and significantly higher ring responses (5 to 8.4 μA), indicating a strong predominance of the 2-electron mechanism with electrogeneration of $H_2O_2$ (**Figure 8c**). This behavior is consistent with the intrinsic inertness of carbonaceous supports, in which oxygen is only weakly adsorbed and preferentially reduced to peroxide.

In contrast, 3% $Fe_3O_4$NO/C (-87.0 to -411.8 μA) displayed markedly higher disk currents and comparatively lower ring currents (1.8 to 4.7 μA), particularly at higher rotation speeds, demonstrating faster ORR kinetics and partial reduction of the electrogenerated peroxide at the catalyst surface. The more pronounced limiting-current plateau (~3.9 μA and -0.37 V) in 3% $Fe_3O_4$NO/C further highlights its enhanced $O_2$-activation capability.

Quantitative analysis of the RRDE data in the potential window from –0.3 to –0.9 V showed that 3% $Fe_3O_4$NO/C provided a 2-fold increase for $H_2O_2$ selectivity in comparison with Vulcan XC72 (**Figure 8d**).

Taken together, the RRDE profiles, electron-transfer numbers, and peroxide yields confirm that Vulcan is highly selective for $H_2O_2$ electrosynthesis, whereas 3% $Fe_3O_4$NO/C improves ORR kinetics and partially suppresses $H_2O_2$ formation through enhanced catalytic reduction pathways.

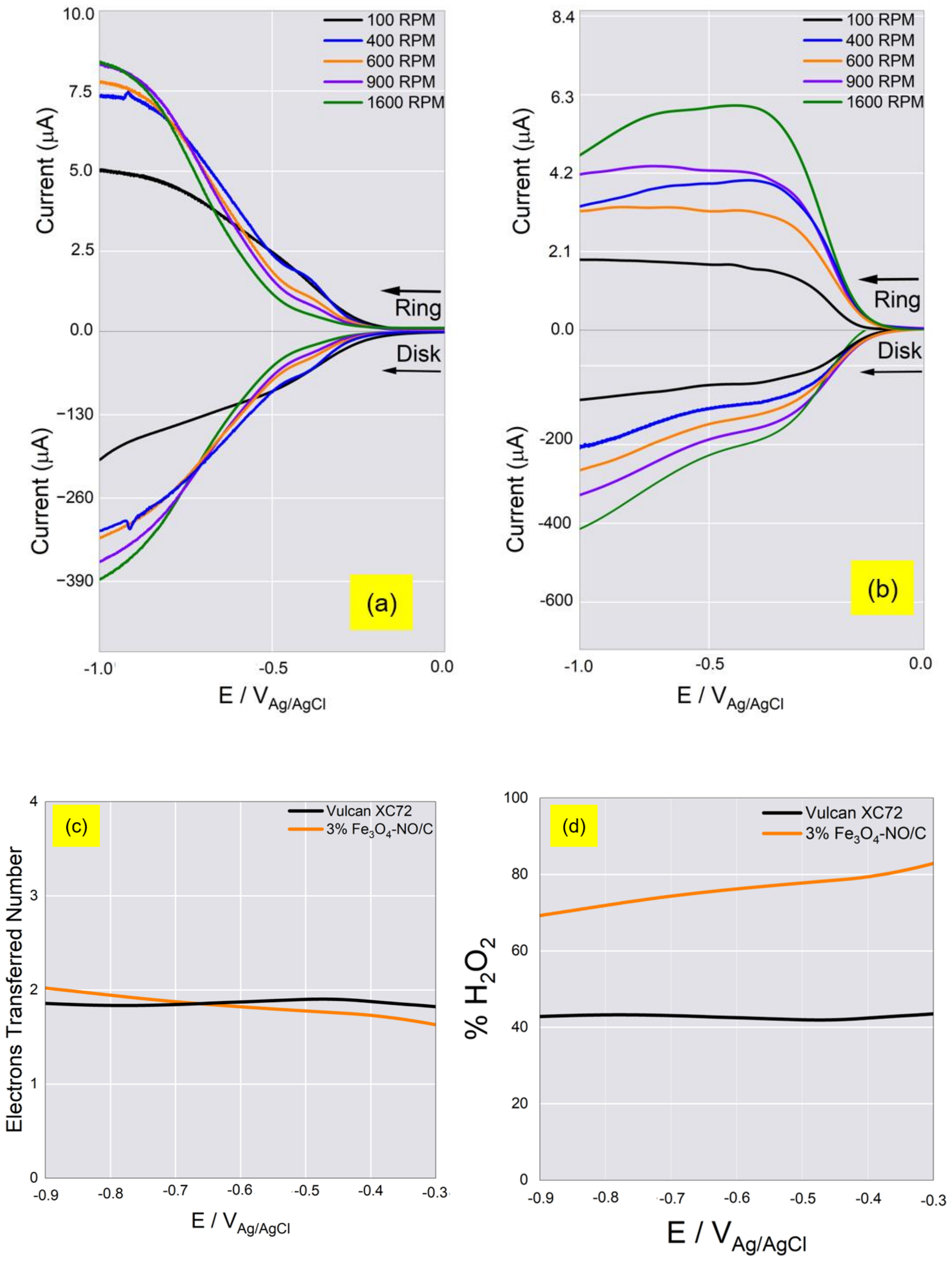

**Figure 8**. LSV curves of Vulcan XC72 (a) and 3% Fe3O4-NO/C (b) on RRDE in $O_2$-saturated 0.1 mol $L^{-1}$ $Na_2SO_4$ (pH = 6.0). Calculated electron transfer number (c) and $H_2O_2$ selectivity (d) through LSV curves.

Barros *et al.* [22] investigated applying $Fe_3O_4$ nanoparticles in ORR supported on Printex and graphene in alkaline medium. Electrochemical studies conducted in 1 mol $L^{-1}$ KOH showed that all tested materials exhibited ORR activity, with $Fe_3O_4$/graphene displaying the highest electrocatalytic performance. This composite exhibited a more positive onset potential and higher current density (1.12 mA $cm^{-2}$ at −0.3 V vs. SCE), outperforming both the $Fe_3O_4$/Printex catalyst (0.38 mA $cm^{-2}$) and the individual components (graphene: 0.85 mA $cm^{-2}$; Printex: 0.29 mA $cm^{-2}$). Rotating ring-disk electrode (RRDE) experiments revealed that the number of electrons transferred during ORR was approximately 2.7 for both composites, indicating a mixed $2e^-/4e^-$ process. The selectivity towards $H_2O_2$ generation reached 68% for $Fe_3O_4$/Printex and 62% for $Fe_3O_4$/graphene at −0.6 V, demonstrating the effective promotion of the two-electron pathway. Additionally, Barros *et al.* [22] conducted long-term stability tests performed via chronoamperometry at −0.3 V for 5.5 hours confirmed the high durability of both catalysts. Notably, $Fe_3O_4$/graphene maintained and slightly improved its performance over time, while $Fe_3O_4$/Printex also showed stable operation with higher $H_2O_2$ selectivity.

#### 3.2.3 Performance of 3% $Fe_3O_4$-NO/C-GDE

GDE 2, prepared with 3% $Fe_3O_4$-NO, was applied in the electrogeneration system to assess the feasibility of the models developed in the previous topic (item 3.2) and to provide an initial evaluation of GDE 2's electrocatalytic performance. **Figure 9** presents the monitoring of [$H_2O_2$] during 60 min of electrolysis conducted in the best range of operational conditions ($j$ = 137 mA $cm^{-2}$; pH = 6.0; $[Na_2SO_4]_0$ = 1 mol $L^{-1}$) with GDE 1 and GDE 2.

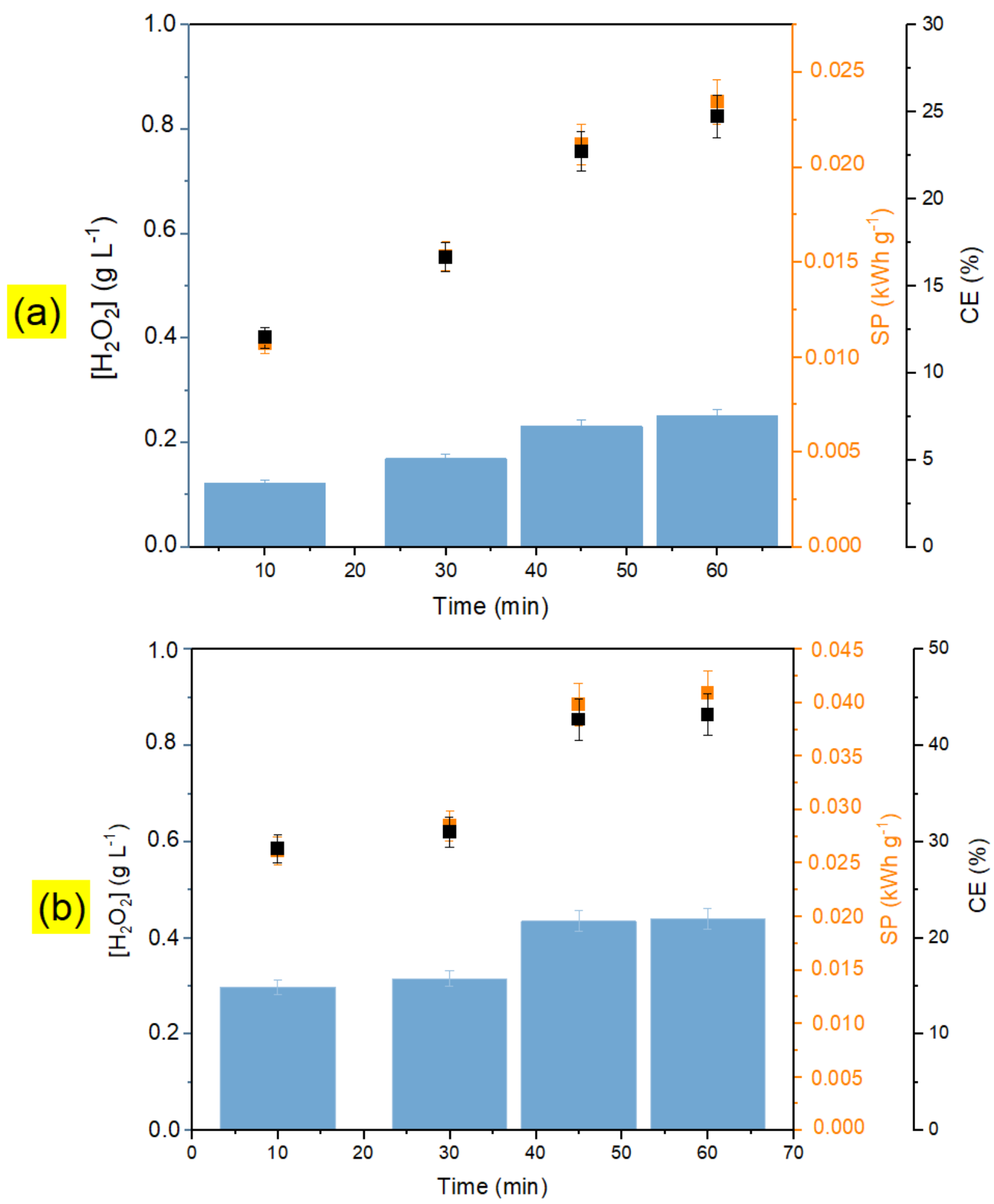

**Figure 9**. Evolution of [$H_2O_2$] electrogenerated applied with GDE 1 (a) and GDE 2 (b). Electrolysis time: 60 min. Experimental conditions: $j$ = 137.5 mA $cm^{-2}$; pH = 6.0; [$Na_2SO_4$]$_0$ = 1 mol $L^{-1}$; anode = platinum.

**Figure 9** highlights the improvement that magnetite nano-octahedral provided for hydrogen peroxide electrogeneration, achieving 91.7 ± 0.14% of $H_2O_2$ electrogenerated ([$H_2O_2$]$_{60}$ = 0.44 ± 0.02 g $L^{-1}$), 2.2% higher than GDE 1 (89.5%, [$H_2O_2$]$_{60}$ = 0.25 g $L^{-1}$). Regarding energetic parameters, the addition of magnetite nanoparticles resulted in a 1.2-fold decrease in SP (GDE 1: 0.010 ± 0.002 kWh $g^{-1}$, GDE 2: 0.012 ± 0.009 kWh $g^{-1}$) and a 1.1-fold increase in CE (GDE 1: 24.7 ± 0.14, GDE 2: 43.1 ± 0.23%).

**Table 2** presents the reported data from previous studies on the use of GDEs made from pure carbonaceous materials, such as Vulcan XC72 and Printex L6. Based on the presented evidence, the modification of Vulcan XC72 with 3% $Fe_3O_4$-NO/C resulted in a 2-fold increase in [$H_2O_2$] yield and a 2-fold decrease in specific energy consumption at the lowest time and current density compared with other studies. Additionally, **Table 2** highlights the main operation conditions applied in some studies with GDE, in bath and flow reactor configurations. This aspect affects the $H_2O_2$ yield but has not been investigated in the present work. This will be addressed in future studies due to the positive effect of solution circulation on process efficiency and energetic parameters [15].

A few articles have presented this type of investigation and employed a distinct approach regarding the use of $Fe_3O_4$ nanoparticles: Zhang *et al.* [54]developed an integrated electro-Fenton system in which $Fe_3O_4$ nanoparticles were immobilized directly onto a rotating gas-diffusion cathode ($Fe_3O_4$/GDE), thereby coupling oxygen diffusion, *in-situ* $H_2O_2$ electrogeneration, and immediate Fenton activation within the same electrode assembly. Under optimized conditions—namely, a cathodic potential of ~0.4 V versus Ag/AgCl at pH 3 and a rotation speed of ~1,500 rpm—the system achieved complete degradation of 50 mg $L^{-1}$ tetracycline within 120 minutes, with a total organic carbon (TOC) removal efficiency of 56.7%. This represented a 1.7-fold increase in degradation rate compared to an equivalent amount of dispersed $Fe_3O_4$ catalyst under identical electro-Fenton conditions. Importantly, leached iron contributions were negligible, confirming that the immobilized Fe3O4 was primarily responsible for electrogenerating and activating $H_2O_2$. Additionally, intermediate-ion chromatography analysis identified hydroxyl radicals as the dominant reactive species, and a tetracycline degradation pathway was proposed based on the detected fragment ions.

**Table 2**. Brief overview of GDE applications for *in-situ* $H_2O_2$ electrogeneration

| GDE | Operational conditions | | | Monitoring parameters | | | |
|---|---|---|---|---|---|---|---|
| | Reactor type | $j$ (mA cm$^{-2}$) | Electrolysis time (min) | [$H_2O_2$] (mg min$^{-1}$ cm$^{-2}$) | SP (kWh kg$^{-1}$) | CE (%) | Reference |
| **Vulcan XC72** | **Batch and undivided cell** | **137.5** | **60** | **0.93** | **40.86** | **24.7** | **This work** |
| **3% $Fe_3O_4$-NO/C** | **Batch and undivided cell** | **137.5** | **60** | **2.10** | **23.43** | **43.1** | **This work** |
| 1% $WO_3$/Vulcan XC72 | Batch and undivided cell | 100.0 | 120 | 1.9 | 17.0 | 62.0 | [42] |
| Printex L6 (PL6C) | Flow-by divided cell | 150.0 | 90 | 1.08 | 24.0 | 55.0 | [36] |
| Printex L6 (PL6C) | Flow-by undivided cell | 50.0 | 120.0 | 0.06 | 105.0 | 25.0 | [38] |
| Printex L6 (PL6C) | Flow-by undivided cell | 125.0 | 120.0 | 1.04 | 0.04 | 80.0 | [15] |

The improved electrochemical performance of the 3% $Fe_3O_4$-NO/C electrode under optimal conditions is closely linked to its structural and surface features. The nano-octahedral shape of $Fe_3O_4$ offers clear crystallographic facets that enhance oxygen adsorption and activation. Even dispersion of the conductive Vulcan XC72 matrix ensures effective electron flow across the electrode. XPS analysis shows an increase in surface hydroxyl groups and oxygen-related functionalities after adding $Fe_3O_4$, which, along with a lower contact angle, improves wettability and allows better electrolyte infiltration into the catalytic layer. These traits help form a robust triple-phase boundary in the gas diffusion electrode, boosting oxygen mass transport and the efficiency of the two-electron ORR process. RRDE measurements confirm that these structural and interface features lead to higher $H_2O_2$ selectivity and altered electron-transfer pathways compared to the carbon support alone, illustrating how optimized operation conditions work synergistically with the catalyst structure rather than independently.

### 3.2.4 GDE 2 and combined process on progestin electrodegradation

The primary objective of this section was to assess the feasibility of combining processes and pathways to generate hydroxyl radicals and degrade both progestins. In this context, **Table 3** presents the removal percentage data for LNG, GES, and TOC for each process combination. Additionally, **Figure S4** shows the concentration decay of LNG and GES during 60 minutes of electrolysis in electrochemical-based processes, including anodic oxidation, anodic oxidation with additional solar photolysis, electro-Fenton, and solar photo-electro-Fenton. All experiments were conducted with GDE 2 as the cathode.

**Table 3**. Experimental conditions and LNG, GES, and TOC removal percentage. Electrochemical cell parameters: $j$ = 137,5 mA cm$^{-2}$; pH = 6.0 and $[Na_2SO_4]_0$ = 1.0 mol L$^{-1}$; Cathode – GDE 2 (3% $Fe_3O_4$-NO/C). $[LNG]_0$ = 1.0 ± 0.52 mg L$^{-1}$ and $[GES]_0$ = 1.0 ± 0.8 mg L$^{-1}$. Electrolysis time = 60 min.

| Experiment | Anode | Solar light | $Fe^{2+}$ (mmol L$^{-1}$) | $LNG_{60'}$ (%) | $GES_{60'}$ (%) | $TOC_{60'}$ (%) | Based degradation process |
|---|---|---|---|---|---|---|---|
| **I** | Pt | N.A. | N.A. | 42.6 | 48.8 | 35.39 | $H_2O_2$ |
| **II** | DSA | N.A. | N.A. | 43.9 | 69.8 | 4.75 | $H_2O_2$ + AO (active anode) |
| **III** | BDD | N.A. | N.A. | 38.6 | 62.0 | 33.44 | $H_2O_2$ + AO (inactive anode) |
| **IV** | Pt | POWER | N.A. | 53.6 | 76.7 | 18.50 | $H_2O_2$ + Solar photolysis |
| **V** | Pt | N.A. | 0.5 | 36.5 | 79.6 | 33.44 | Electro-Fenton |
| **VI** | Pt | POWER | 0.5 | 45.0 | 92.4 | 11.08 | Solar Photo Electro-Fenton |
| **VII** | BDD | N.A. | 0.5 | 63.5 | 92.9 | 38.16 | Electro-Fenton + AO (inactive anode) |
| **VIII** | BDD | POWER | 0.5 | 72.6 | 100 | 60.50 | Solar Photo Electro-Fenton + AO |

N.A. = not applicable.

In Experiment I, where a Pt anode was employed in the absence of iron and solar irradiation, only moderate removals were obtained for LNG (42.6%), GES (48.8%), and TOC (35.4%). This performance can be attributed mainly to direct oxidation by electrogenerated $H_2O_2$, whose limited production and instability under dark conditions reduce the oxidative capacity of the system [55]. It is essential to highlight that Experiment I experimental conditions (**Table 1 – runs 9 to 10**) were capable of electro-synthesizing 95.4 ± 0.44 % of $H_2O_2$ ($[H_2O_2]_{60'}$ = 0.2517 ± 0.02 g $L^{-1}$) but without providing high progestins and TOC removals.

When replacing Pt with DSA (Experiment II), a significant improvement was observed for GES degradation (69.8%), but only a marginal improvement for LNG (43.9%), accompanied by poor mineralization efficiency (4.75%). This behavior indicates the role of the "active" anode in promoting surface-bound hydroxyl radicals (•OH_ads), which exhibit high reactivity but low mineralization capability due to their surface confinement [56].

The use of BDD (Experiment III) as an "inactive" anode enhanced mineralization (33.4%), while GES degradation (62.0%) was slightly lower than with DSA. BDD is well-known for generating free •OH from water discharge with higher oxidation potential, thus promoting more extensive mineralization of organics [56,57].

The addition of solar irradiation in Experiment IV (Pt/POWER) markedly increased the degradation efficiency for both LNG (53.6%) and GES (76.7%), but mineralization remained moderate (18.5%). These results highlight the beneficial role of solar photolysis in accelerating $H_2O_2$ decomposition and promoting hydroxyl radical generation, thereby boosting substrate degradation, though not completely enhancing mineralization [58].

The introduction of $Fe^{2+}$ ions (0.5 mmol $L^{-1}$) in Experiments V–VIII significantly enhanced the oxidative performance due to the occurrence of the Electro-Fenton (EF) reaction. In the absence of solar light, Pt/EF (Experiment V) promoted moderate degradation (LNG: 36.5%, GES: 79.6%) and TOC removal (33.4%), reflecting the efficiency of homogeneous •OH production in solution. Interestingly, Solar PEF with Pt (Experiment VI) provided superior GES degradation (92.4%) but limited mineralization (11.1%), suggesting that while the photo-electro Fenton effect increased the oxidation of the parent molecule, partial oxidation products persisted in solution [59–61].

When employing BDD as anode (Experiment VII) under EF conditions, remarkable improvements were recorded for both LNG (63.5%) and GES (92.9%), along with 38.1% TOC removal. The synergistic contribution of homogeneous •OH (Fenton reaction) and heterogeneous •OH (BDD surface) accounts for the higher mineralization efficiency observed [59–61].

Finally, Solar PEF coupled with BDD (Experiment VIII) achieved the best overall performance, with complete GES degradation (100%), high LNG removal (72.6%), and the highest TOC mineralization (60.5%). These findings demonstrate the strong synergy between electro-Fenton chemistry, photo-Fenton enhancement, and the powerful oxidizing ability of BDD electrodes, resulting in near-complete mineralization. Such synergistic behavior has been widely reported, especially for persistent organic pollutants, where solar irradiation accelerates $Fe^{3+}/Fe^{2+}$ cycling and enhances hydroxyl radical availability [62–64].

**Figure 10** presents the cumulative [$H_2O_2$] from combined electrochemical and photochemical experiments, plotted against SP and CE factors—all experiments were conducted with GDE 2 as the cathode.

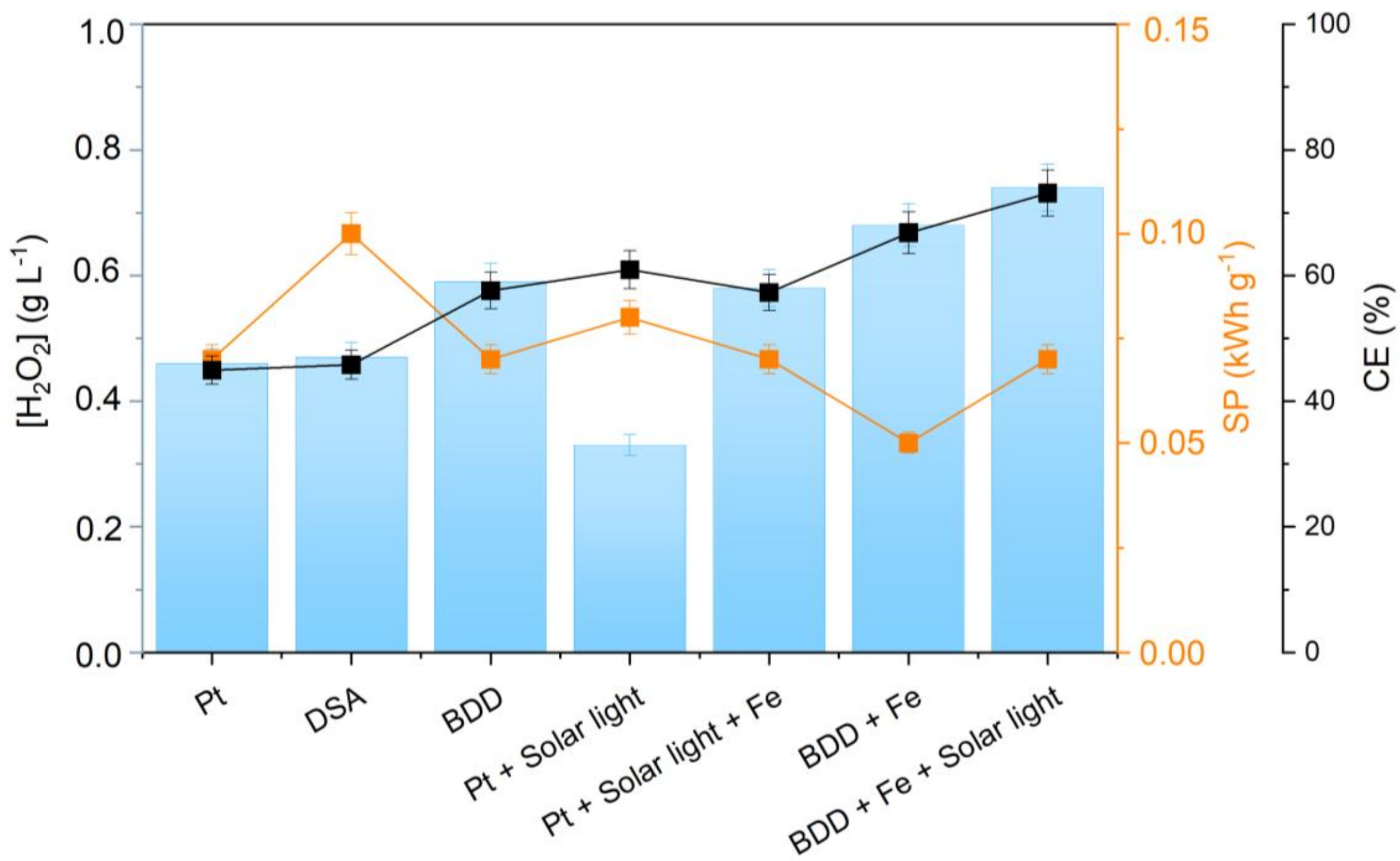

**Figure 10 -** Energetic parameters monitoring in GDE 2 combined electrochemical and Solar-photochemical process. Experimental conditions: $j$ = 137.5 mA cm$^{-2}$; pH = 6.0; $[Na_2SO_4]_0$ = 1 mol L$^{-1}$; $[LNG]_0$ = 1.0 ± 0.52 mg L$^{-1}$ and $[GES]_0$ = 1.0 ± 0.8 mg L$^{-1}$. Electrolysis time: 60 min.

Based on **Figure 10**, we can see that Pt- and DSA-based systems showed moderate $H_2O_2$ generation (~0.45 g L$^{-1}$), with DSA exhibiting higher SP, confirming its lower energy efficiency due to surface-confined •OH. In contrast, BDD favored higher $H_2O_2$ accumulation (~0.58 g L$^{-1}$) and lower SP, consistent with its greater mineralization capacity.

The use of solar light with Pt reduced the $H_2O_2$ concentration (~0.30 g L$^{-1}$) via photolytic decomposition, accelerating progestin degradation but lowering CE. The addition of $Fe^{2+}$ in electro-Fenton systems improved $H_2O_2$ utilization and decreased SP, particularly with Pt + Solar light + Fe, although mineralization remained limited by the persistence of intermediates.

The best performance was achieved with BDD + Fe, especially under solar irradiation, which resulted in the highest $H_2O_2$ accumulation (~0.73 g $L^{-1}$), a CE above 70%, and a relatively low SP (~0.06 kWh $g^{-1}$). This correlates with the superior mineralization efficiency observed in **Table 3**, confirming the synergistic effect of solar-assisted electro-Fenton with BDD.

The reusability assay was applied to evaluate the stability and efficiency of GDE 2 in generating $H_2O_2$ electrolytes and degrading progestins. **Figure 11a** shows the GDE 2 performance for three 60-minute cycles, where [$H_2O_2$], SP, and CE achieved stable values of 0.75 ± 0.04 g $L^{-1}$, 0.07 ± 0.003 kWh $g^{-1}$, and 73.7 ± 3.9%, respectively. These results confirm the robustness of the gas diffusion electrode (GDE) and the electro-Fenton system, with no significant loss of efficiency upon reuse. The stable CE and moderate SP values indicate good operational sustainability, consistent with reports on the long-term reliability of BDD-based EF systems [65].

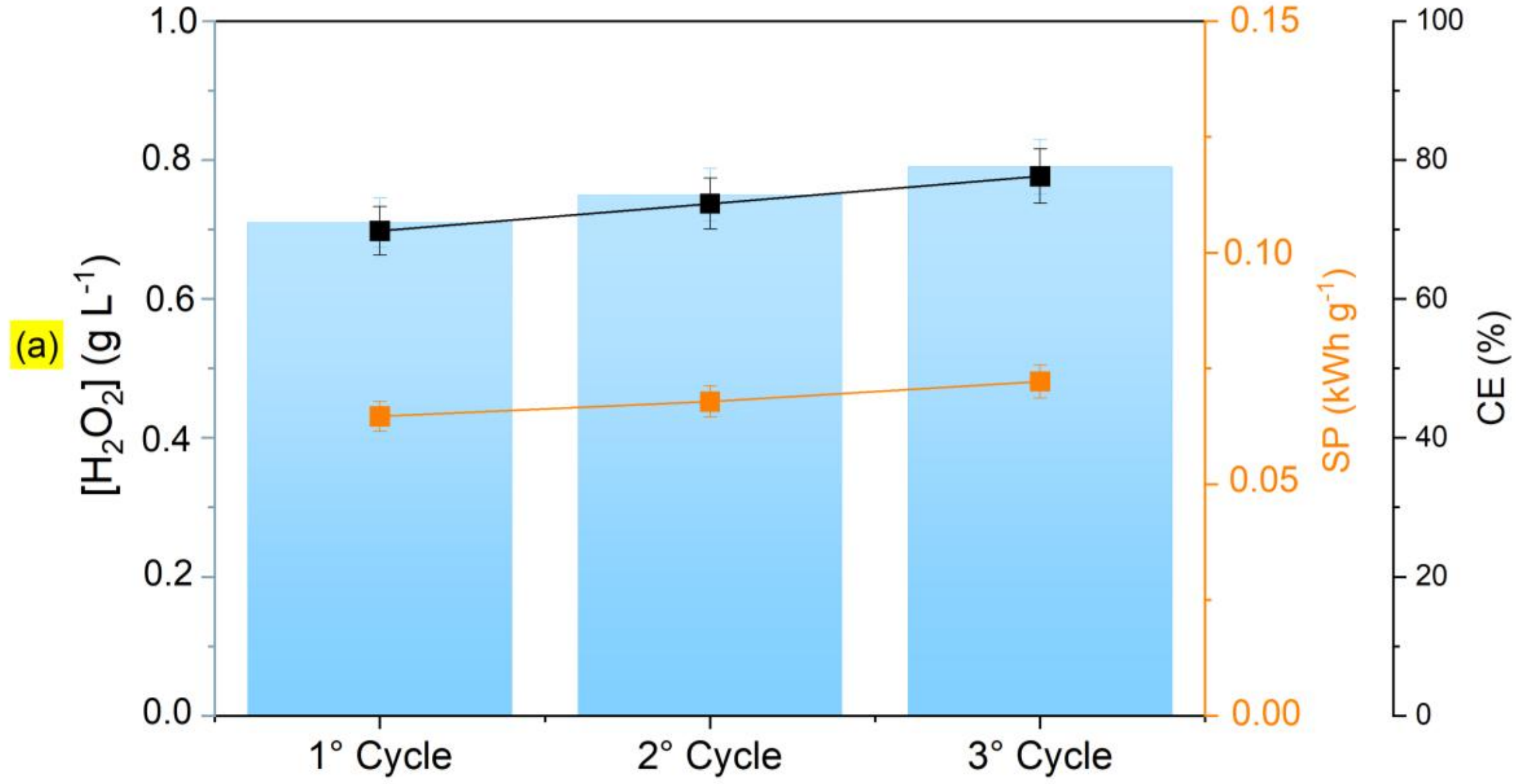

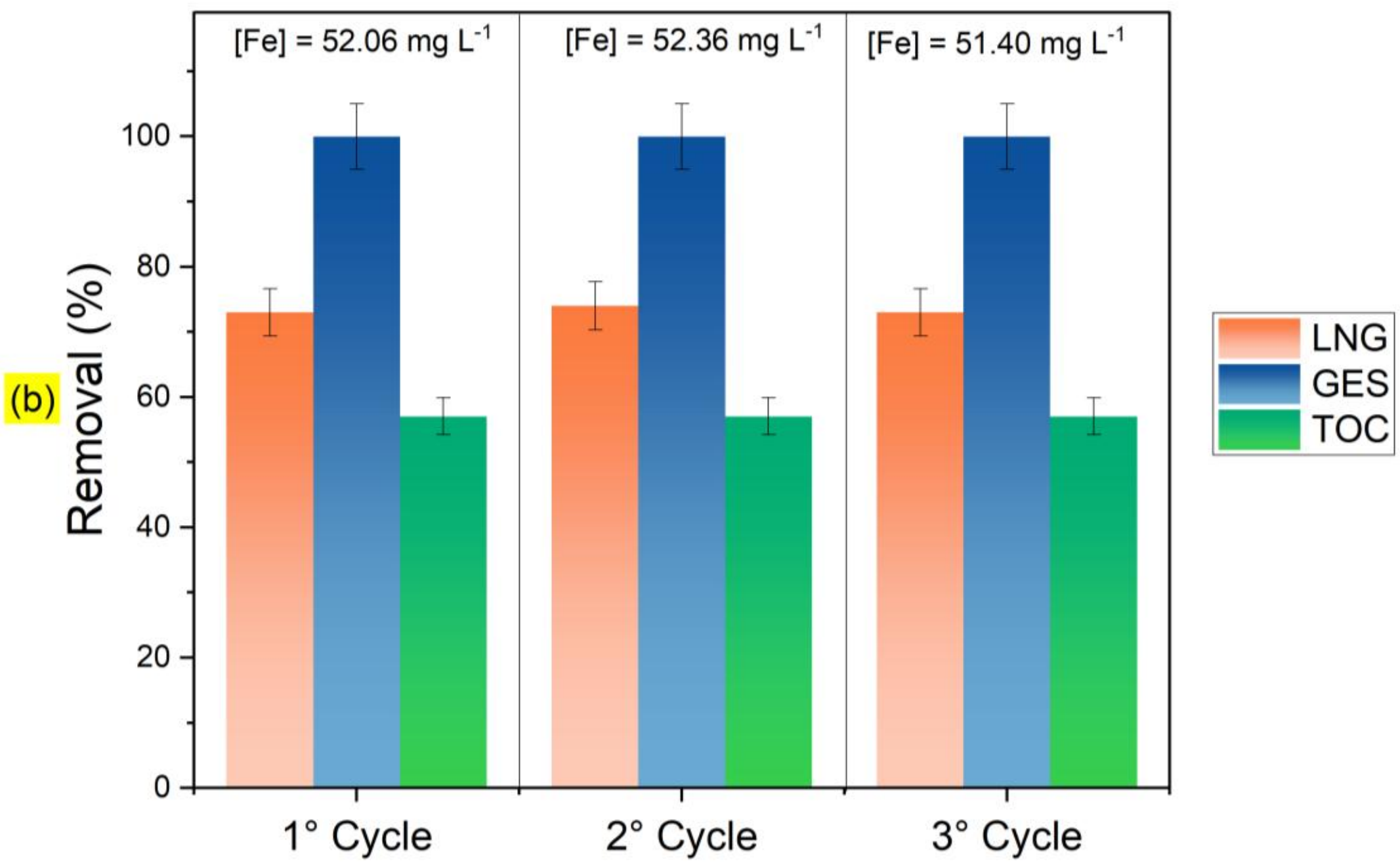

**Figure 11** - System reusability in three consecutive Solar-Photo-Electro Fenton for progestins degradation: (a) $H_2O_2$ production and energetic parameters; (b) LNG, GES, TOC, and total iron monitoring. Experimental conditions: $j$ = 137.5 mA cm$^{-2}$; pH = 6.0; $[Na_2SO_4]_0$ = 1 mol L$^{-1}$; $[LNG]_0$ = 1.0 ± 0.52 mg L$^{-1}$ and $[GES]_0$ = 1.0 ± 0.8 mg L$^{-1}$. Electrolysis time: 60 min.

**Figure 11(b)** presents the removal of LNG, GES, and TOC over the three cycles. GES degradation remained consistently high (≈approximately 95–97%), whereas LNG removal stabilized at around 70–72%. TOC mineralization was lower (≈approximately 50–55%) but remained stable across all cycles, reflecting the persistence of oxidation intermediates. The GDE stability during the electrochemical process is further supported by the nearly constant residual Fe concentration (51–52 mg L$^{-1}$), suggesting minimal iron loss and practical $Fe^{2+}/Fe^{3+}$ cycling. Similar behavior has been reported in electro-Fenton treatments of pharmaceuticals, where catalytic stability ensures reproducible performance [55].

In addition to optimizing electrochemical performance, the stability of the $Fe_3O_4$-NO/C electrode under the tested operating conditions was thoroughly examined. Changes in pH and the use of relatively high current densities could potentially cause phase changes, surface restructuring, or mechanical breakdown in oxide catalysts. Still, the current results show that the nano-octahedral $Fe_3O_4$ structure remains stable across the tested parameters. Reusability tests over three consecutive Solar Electro-Fenton cycles demonstrated consistent pollutant removal and $H_2O_2$ production, indicating the catalyst's retained activity. ICP-MS analysis of iron leaching showed minimal Fe release, confirming the structural stability and secure anchoring of $Fe_3O_4$ on Vulcan XC72. These results confirm that the optimized electrochemical conditions do not negatively impact the catalyst's reactivity or mechanical stability, supporting its use for repeated, long-term operation.

Overall, the cycling experiments confirm that the solar-assisted EF system, incorporating BDD and GDE, not only achieves high degradation and mineralization of progestins but also exhibits excellent stability and reusability, thereby reinforcing its applicability for continuous or semi-continuous wastewater treatment.

# 4. CONCLUSIONS

This study demonstrated the successful synthesis and application of $Fe_3O_4$-NO/C composites for *in situ* electrogeneration of $H_2O_2$ using GDE. The materials exhibited favorable physicochemical properties, including high surface area, enhanced hydrophilicity, and strong nanoparticle–carbon interactions.

Statistical optimization revealed that the current density, pH, and concentration of the supporting electrolyte significantly affect electrogeneration performance as a function of the applied response. Hydrogen peroxide electrosynthesis was monitored using $H_2O_2$ yield, CE, and SP as independent variables, enabling us to evaluate the process with robust experimental data. The primary evidence highlighted the feasibility of producing H2O2 from carbonaceous materials. However, high yields are not directly related to high current efficiency or high specific production, as confirmed by PCA data.

Optimizing antagonizing factors, as conducted in PCA, provides further insights into the multivariate influences. The evidence highlights the relationship between the principal components (PC1 = CE and PC2 = SP) and the loadings. It suggests that the level of each variable should be adjusted to reduce specific production and increase current efficiency. As a result, the data allowed us to estimate the best range of operational parameters ($[Na_2SO_4]_0$ = 1.0 mol $L^{-1}$, pH = 6.0, and $j$ = 137.5 mA $cm^{-2}$).

In this sense, the optimized condition was applied to the 3% $Fe_3O_4$-NO/C-GDE, achieving notable improvements in current efficiency and specific production compared to Vulcan XC72 (SP: 0.012 ± 0.009 kWh $g^{-1}$; CE: 43.1 ± 0.23%). Furthermore, the composite electrode showed excellent performance in degrading endocrine-disrupting progestins (LNG and GES) under solar-assisted electro-Fenton conditions, with high removal efficiency (70 and 100%) and operational stability across multiple cycles ($[H_2O_2]$ = 0.75 ± 0.04 g $L^{-1}$, SP = 0.07 ± 0.003 kWh $g^{-1}$, and CE 73.7 ± 3.9%) without GDE leaching. These results highlight the potential of 3% $Fe_3O_4$-NO/C-GDEs for scalable, low-cost EAOP applications to remove micropollutants in wastewater treatment.

**Credit authorship contribution statement**

**Conceptualization,** J.M.S.J.; **methodology,** J.M.S.J., C.O.C, C.C.A, B.L.B., J.P.C.M., and A.B.T.**; software,** J.M.S.J., J.P.C.M., and A.B.T.**; validation,** J.M.S.J.**; formal analysis,** J.M.S.J., C.C.A, J.P.C.M., and A.B.T.**; investigation** J.M.S.J., C.O.C, C.C.A, B.L.B., J.P.C.M., and A.B.T.**; resources** M.C.S.; **data curation,** J.M.S.J., C.C.A, J.P.C.M., and A.B.T.**; writing-original draft preparation,** J.M.S.J., C.C.A., B.L.B., J.P.C.M., and A.B.T.**; writing - review and editing,** J.M.S.J., B.L.B., and M.C.S.; **visualization,** J.M.S.J., and M.C.S.; **supervision,** M.C.S.; **project administration,** J.M.S.J., and M.C.S.; **funding acquisition,** M.C.S. All authors have read and agreed to the published version of the manuscript.

**Data availability**

The raw/processed data required to reproduce these findings cannot be shared for legal or ethical reasons.

**Declaration of competing interests**

The authors declare that they have no known competing financial interests or personal relationships that could have influenced the work reported in this paper.

**Acknowledgments**

The authors thank the Research Group in Advanced Oxidation Process (AdOx) of the Polytechnic School of São Paulo University (POLI-USP) for its support and access to chromatography analysis. The authors would like to thank Fundação de Amparo à Pesquisa do Estado de São Paulo (FAPESP, #2021/05364-7, #2021/14394-7, #2022/10484-4, #2022/12895-1, #2022/15252-4, #2024/11134-2) for the financial support. The authors are also grateful for Coordenação de Aperfeiçoamento de Pessoal de Nível Superior (CAPES) and Conselho Nacional de Desenvolvimento Científico e Tecnológico (CNPq) (#303943/2021-1, #308663/2023–3, #402609/2023–9) for their support.